\documentclass[preprints,article,accept,moreauthors]{Definitions/mdpi} 

\firstpage{1} 
\pubvolume{xx}
\issuenum{1}
\articlenumber{5}
\pubyear{2020}
\copyrightyear{2020}
\usepackage{subfigure}
\history{Received: date; Accepted: date; Published: date}

\Title{Acoustic streaming generated by sharp edges: the coupled influences of liquid viscosity and acoustic frequency}

\Author{Chuanyu Zhang $^{1,}$*, Xiaofeng Guo $^{1,3}$, Laurent Royon $^{1}$ and Philippe Brunet $^{2,}$*}

\AuthorNames{Chuanyu Zhang, Xiaofeng Guo, Laurent Royon and Philippe Brunet}

\address{%
$^{1}$ \quad Universit\'e de Paris, Laboratoire Interdisciplinaire des Energies de Demain, UMR 8236 CNRS, F-75013 Paris, France; chuanyu.dream@gmail.com\\
$^{2}$ \quad Universit\'e de Paris, Laboratoire Mati\`ere et Syst\`emes Complexes, UMR 7057 CNRS, F-75013 Paris, France; philippe.brunet@univ-paris-diderot.fr \\
$^{3}$ \quad Universit\'e Gustave Eiffel, ESIEE Paris, F-93162, Noisy le Grand, France}

\corres{Correspondence: chuanyu.dream@gmail.com (C.Z.), philippe.brunet@univ-paris-diderot.fr (P.B.); Tel.: +33-157276272 (P.B.)}

\abstract{Acoustic streaming can be generated around sharp structures, even when the acoustic wavelength is much larger than the vessel size. This sharp-edge streaming can be relatively intense, \textcolor{blue}{owing to the strongly focused inertial effect experienced by the acoustic flow near the tip.} We conducted experiments with Particle Image Velocimetry to quantify this streaming flow through the influence of liquid viscosity $\nu$, from 1 mm$^2$/s to 30 mm$^2$/s, and acoustic frequency $f$ from 500 Hz to 3500 Hz. Both quantities supposedly influence the thickness of the viscous boundary layer $\delta = \left(\frac{\nu}{\pi f}\right)^{1/2}$. For all situations, the streaming flow appears as a main central jet from the tip, generating two lateral vortices beside the tip and outside the boundary layer. As a characteristic streaming velocity, the maximal velocity is located at a distance of $\delta$ from the tip, and it increases as the square of the acoustic velocity. We then provide empirical scaling laws to quantify the influence of $\nu$ and $f$ on the streaming velocity. Globally, the streaming velocity is dramatically weakened by a higher viscosity, whereas the flow pattern and the disturbance distance remain similar regardless of viscosity. Besides viscosity, the frequency also strongly influences the maximal streaming velocity.}


\begin{document}


\section{Introduction}

Acoustic streaming (AS) denotes the steady flow generated by an acoustic field in a fluid. Mathematically, it can be explained by the nonlinear coupling between acoustic wave and hydrodynamic momentum conservation equations. Physically, the underlying mechanism of AS comes from the dissipation of acoustic energy within the fluid, which induces spatial gradient of momentum, and in turn creates a time-averaged effective forcing \cite{Westervelt1953,Nyborg1953,Lighthill1978,Friend2011,Eckart1948,Rayleigh1884,Schlichting,Nyborg1958,Riley1998a,Rayleigh2013}.

The phenomenon has attracted researcher's attention since Faraday's observations in 1831 \cite{Faraday1831}, who reported that light particles on vibrating plates spontaneously form steady clusters. More recently and especially in the context of microfluidics, AS has been proven to be a suitable technique for fluid and particle handling in various situations \cite{Friend2011}.
 \textcolor{blue}{We wish to point out the studies on fluid mixing at low-Reynolds number \cite{Sritharan2006}, particles manipulation and sorting \cite{Franke2010,Lenshof2012,Sadhal2012a,Muller2013,Skov2019,Qiu2019}, particles patterning \cite{Voth2002,Vuillermet2016} or heat transfer \cite{Legay2012,Loh2002}, among others}.

Amongst different sorts of acoustic streaming, the one relevant in microfluidics situations usually involves viscous stress along walls or obstacles, generated by no-slip conditions, and resulting in the presence of a viscous boundary layer (VBL). It is \textcolor{blue}{referred to} \textit{Rayleigh-Schlichting streaming} \cite{Rayleigh1884,Schlichting,Nyborg1958,Riley1998a,Friend2011,Rayleigh2013}, and is different from that induced by acoustic attenuation in the bulk of fluid. The bulk acoustic streaming is denoted as \textit{Eckart streaming}\cite{Eckart1948,Nyborg1953} and becomes significant only with high frequencies ($>$ MHz) or with very viscous liquids, so that the attenuation length is smaller than - or of the same order as - the vessel size \cite{Kamakura1996,Brunet2010,Moudjed2014}. In Rayleigh-Schlichting streaming, a non-zero time-averaged vorticity is generated inside the unsteady VBL \cite{Schlichting} of typical thickness $\delta = \left(\frac{2 \nu}{\omega}\right)^{\frac{1}{2}}$, where $\nu$ is the kinematic viscosity, and $\omega = 2 \pi f$ the acoustic angular frequency. This vorticity appears in the form of an array of eddies pairs \cite{Rayleigh1884,Schlichting,Rayleigh2013}, denoted as inner vortices, along the channel walls \cite{Andrade1931,Muller2013,Valverde2015}. This vorticity extends its influence beyond the VBL and in turn induces larger-scale eddies of width $\lambda /2$ \cite{Andrade1931,Hamilton2002} in the fluid bulk, where $\lambda = \frac{c_s}{f}$ is the acoustic wavelength and $c_s$ the speed of sound. Rayleigh-Schlichting streaming is generally treated within the incompressibility framework. 

Traditional acoustic streaming in microchannels is achieved by adjusting the channel width $w$ and the wavelength $\lambda$ to ensure a resonance condition, typically obtained when $w \simeq \lambda/2$ \cite{Wiklund2012a}. However, recent studies evidenced that relatively intense streaming could be generated by designing microchannels with sharp structures along the walls \cite{Huang2013a,Huang2014,Nama2014,Nama2016a,DoinikovMN2020,Zhang2019,Zhang2020} excited by acoustic waves. The sharp structures can be easily prototyped by the facilities offered by microfabrication in clean rooms, e.g. with photolithography. One of the main advantages of ''sharp-edge streaming'' is that it can be generated at relatively low frequency, typically in a range between a few hundred Hz and several kHz (but it is observed for much higher frequency as well \cite{DoinikovMN2020}). Within this low frequency range, numerous performant and stable piezoelectric transducers are available for low cost, and can be supplied with inexpensive amplifiers. Other advantages of operating at relatively low frequency include : efficient acoustic coupling between the transducer and the solid in contact, and negligible acoustic dissipation within the liquid.
Finally, previous experiments reported that near the tip of the sharp edge, the streaming velocity can be very strong \cite{Huang2013a,Huang2014,Nama2014,Ovchinnikov2014}, and can even be comparable to the vibration velocity, hence up to several hundreds of mm/s \cite{Zhang2019,Zhang2020} at a typical distance $\delta$ from the tip. Benefiting from these strong disturbances within the fluid inside a microchannel, various applications using sharp structures streaming have been developed: mixing processes \cite{Huang2018a,Nama2014}, bio-particle control \cite{Leibacher2015,Cao2016}, as well as various on-chip devices \cite{Huang2014,Bachman2018}. 

The present study aims to investigate the influence of both liquid kinematic viscosity $\nu$ and acoustic frequency $f$, on the streaming flow magnitude and pattern. The interest of this study is based on that one of the key parameters of sharp-edge streaming is the thickness of the VBL, which depends on both $f$ and $\nu$. Actually, three main dimensionless numbers involve $\delta$: the ratio of the tip diameter and $\delta$, $d^*=\frac{2 r_c}{\delta}$, the ratio with respect to the channel depth $p$, $p^* = \frac{p}{\delta}$ and the ratio between the channel width $w$ and $\delta$, $w^*=\frac{w}{\delta}$. Sharp-edge streaming is defined by the \textit{sharpness} condition $d^* < $1 \cite{Ovchinnikov2014}, and almost no streaming could be noticed even at relatively high forcing when $d^* \gg $1 \cite{Zhang2019,Zhang2020}. In the typical framework with water and $f$ of a few kHz (let us say between 2500 and 6000 Hz as in previous studies), $\delta$ ranges between 7.3 and 11.3 $\mu$m, so that the two other ratios $w^* , p^* \gg 1$, for microfluidic channels, typically thicker than 50 $\mu$m.

Additionally, quantifying the influence of viscosity distinguishes sharp edge acoustic streaming from classical ones. In classical Rayleigh-Schlichting streaming, the flow is found to be independent on viscosity providing that the VBL thickness $\delta$ is much thinner than the vessel size \cite{Nyborg1958,Riley1998a,Costalonga2015}. For sharp-edge streaming in microchannels or in wider vessels, it is found that this independence on viscosity is lost even if $\delta$ remains thin compared to the channel width $w$ or depth $p$ \cite{Ovchinnikov2014}. Ovchinnikov \textit{et al.}'s perturbative theory predicts a decrease of the typical streaming velocity $V_s$ with $\nu$, though with a subtle dependence on the sharp-edge geometry. With a viscous enough liquid and/or a low enough frequency, the dimensionless lengths $p^*$ or $w^*$ can fall into the order of one. Under this condition, an overlap between geometrical confinement and the intrinsic nature of sharp-edge streaming makes it more complex to determine the influence of $\nu$ and $f$ on the flow. On this latter point, equation (22) from \cite{Ovchinnikov2014} predicted a typical streaming velocity in cylindrical coordinate $(r,\phi)$ as :

\begin{equation}
    V_s(r)=\frac{V_a^2}{\nu}\frac{\delta^{2n-1}}{h^{2n-2}}H_{\alpha}(\frac{r}{\delta})
    \label{eq:ovchinnikov_vs}
\end{equation} 

\noindent where $V_a$ is the amplitude of the acoustic velocity, $n$ is a coefficient that depends on $\alpha$ as $n = \frac{\pi}{2 \pi - \alpha}$; $h$ is the length scale of the sharp-edge height. The function $H_{\alpha}(\frac{r}{\delta})$ contains the radial profile of the streaming flow. It is worth noticing that eq.~(\ref{eq:ovchinnikov_vs}), supposedly valid in the range $r_c < \delta$, does not exhibit any dependence on $r_c$.

The present study intends to quantify the coupled role of viscosity and excitation frequency in both the streaming flow pattern and magnitude. The paper is organised as follows: section 2 described the experimental setup and visualisation method. Then in section 3 and 4 are presented respectively the results at different viscosities and different frequencies. Finally, section 5 summarises the main results and conclusions.

\begin{table}
\caption{Definition of the main physical quantities}
\label{table:assembly}
\centering
\small
\renewcommand{\arraystretch}{1.25}
\begin{tabular}{l l}
\hline\hline
\multicolumn{1}{c}{Quantity} &
\multicolumn{1}{c}{Abbreviation} \\
\hline
Kinematic viscosity & $\nu$ \\
Viscous boundary layer thickness & $\delta$ \\
Tip angle of sharp edge & $\alpha$ \\
Height of the sharp edge & $h$ \\
Radius of curvature of the tip & $r_c$ \\
Width of the microchannel & $w$\\
Depth of the microchannel & $p$\\
Acoustic frequency & $f$  \\
Acoustic angular frequency & $\omega$  \\
Amplitude of acoustic displacement & \textbf{A} \\
Amplitude of acoustic velocity & $\textbf{V}_a$\\
Amplitude of acoustic velocity far from the tip & $V_a$\\
Streaming velocity & $\textbf{V}_s$  \\
Maximum streaming velocity & $V_{s max}$\\
Fitting coefficient relating $V_{s max}$ and $V_a^2$ & $\theta$ \\ 
\hline\hline
\end{tabular}
\normalsize
\end{table}

\section{Experimental setup}

\begin{figure}
\centering
\includegraphics[width=\textwidth]{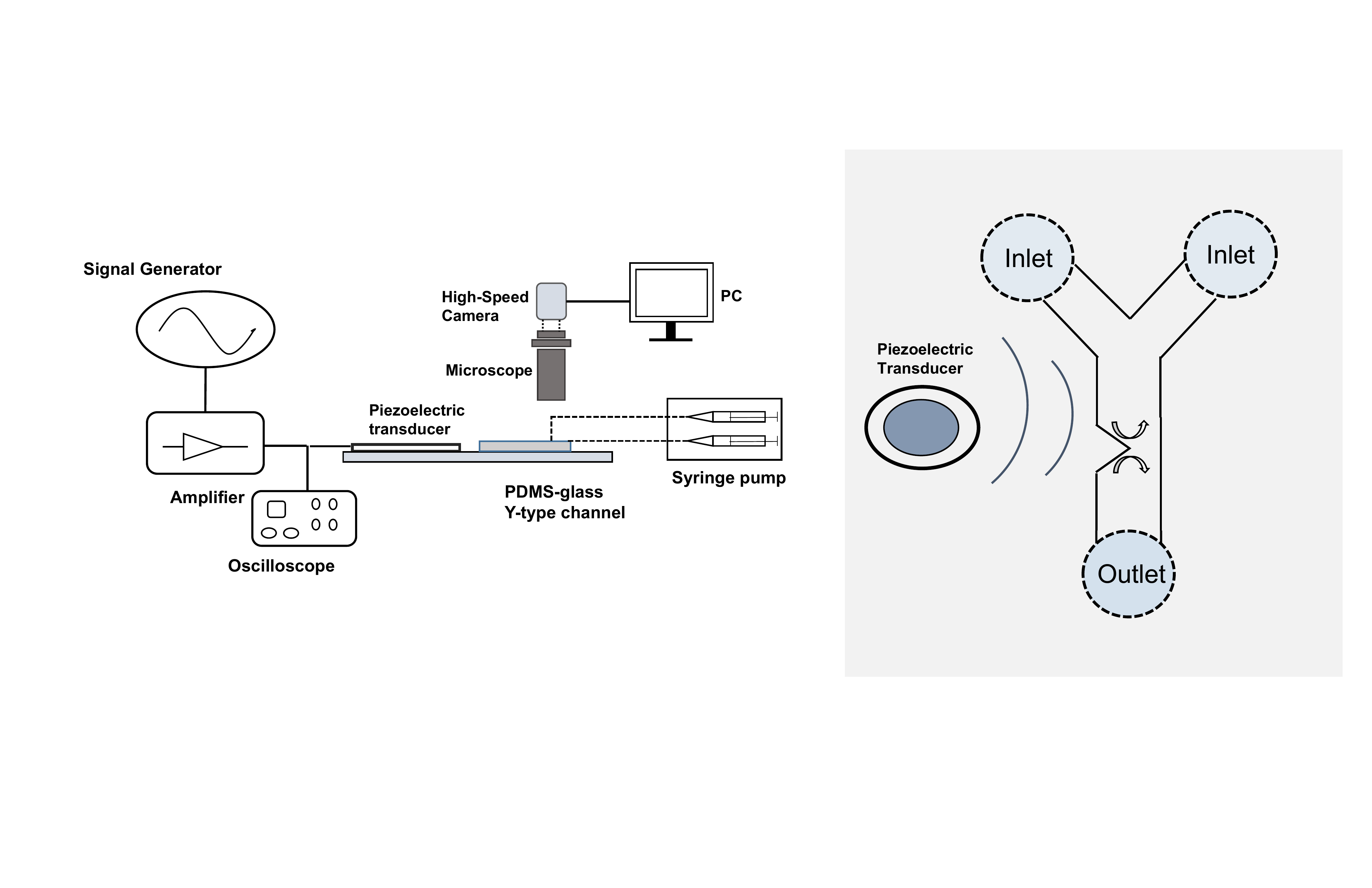}
\caption{\textit{Left -} Sketch of the experimental setup. A piezoelectric transducer is glued on a microscope glass slide, which is used as a coverslip for a PDMS microchannel with one or several sharp-edge structures. The transducer is supplied with a function generator and a home-made amplifier, adjusted by the peak-to-peak voltage monitored with an oscilloscope. The fluid seeded with fluorescent particles is brought by a syringe pump through two inlets. The flow inside the microchannel is visualised by a high-speed camera connected to a binocular microscope. \textit{Right -} The piezo-transducer generates an acoustic wave within the Y-shaped channel. In the vicinity of the sharp-edge structure, the acoustic wave generates a streaming flow.}
\label{fig:setup}
\end{figure}  

\subsection{Microchannel and acoustic wave}

\begin{figure}
\centering
\subfigure[]{\includegraphics[width=0.6\textwidth]{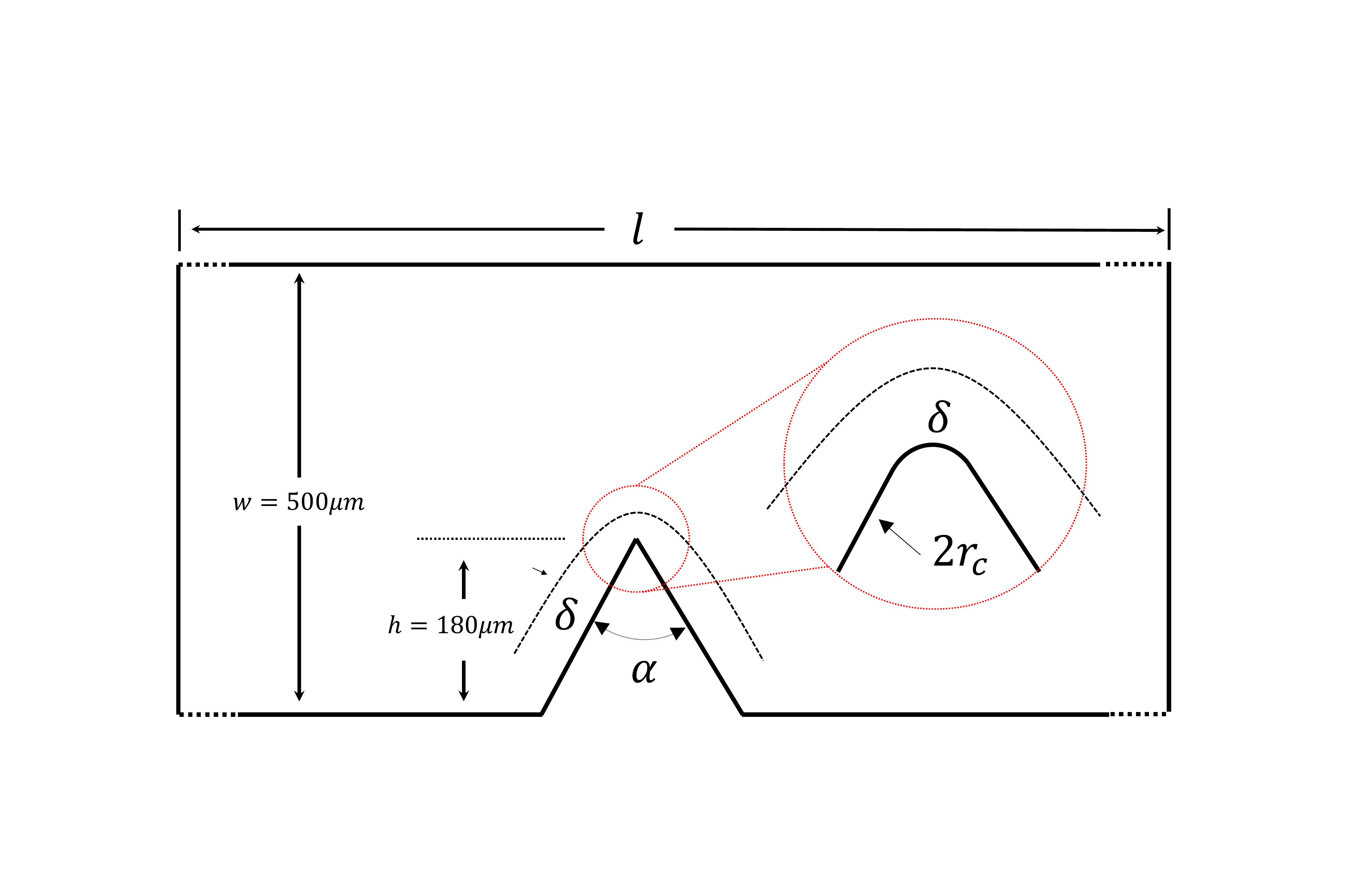}}
\subfigure[]{\includegraphics[width=0.38\textwidth]{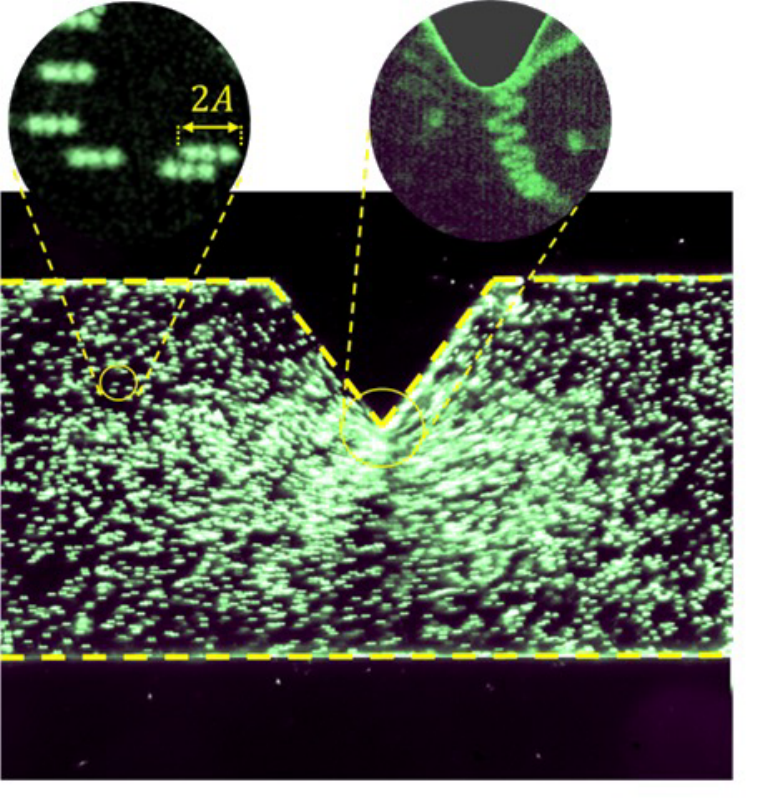}}
\caption{(a) Geometry of the microchannel and sharp-edge. (b) Trajectories of individual particles (diameter 4.9 $\mu$m), over several periods, \textcolor{red}{for the left-hand-side zoom-in image, the frame per second (fps) equals $4 f$ = 10000 fps, while for the right-hand-side one, the fps equals $10 f$ = 25000 fps, the two images have the same exposure time $1/(10f) = 1/25000$s}. Far from the tip, the flow is oscillating at frequency $f$ and amplitude $A$, as testified by the segment described by each particle. Close to the tip, the trajectories of the particles show a superposition of oscillations with higher amplitude due to the sharp edge and advection due to the intense streaming flow.}
\label{fig:flow_va}
\end{figure}

The experimental setup is sketched in Figure \ref{fig:setup}, and presented in more details in \cite{Zhang2019}. It is built around a Y-shaped Polydimethylsiloxane (PDMS) microchannel devised by standard photolithography techniques: starting from a SU8 resist-made mould of thickness 50 $\mu$m made on a silicon wafer, a mixture of PDMS (Sylgard 184) with 10\% in mass of curing agent is poured on the SU8 mould and forms a 2.5-mm-thick layer on top of the wafer. \textcolor{blue}{After a baking at 65 $^{\circ}$C for 4 hours,} the PDMS mixture is then sealed and \textcolor{blue}{attached to a glass coverslip after a 1 mn O$_2$ plasma treatment of both faces}. A PDMS microchannel of depth $p$ = 50 $\mu$m is then obtained. The width $w$ is equal to 500 $\mu$m. Its geometrical dimensions are detailed in Figure \ref{fig:flow_va}-(a). Sharp edges with different angles (30$^{\circ}$, 60$^{\circ}$, 80$^{\circ}$, and 90$^{\circ}$) could be fabricated from various moulds, and previous studies evidenced that a sharper tip and more acute angle would lead to stronger streaming under the same forcing amplitude \cite{Huang2013a,Huang2014,Nama2014,Zhang2019,Zhang2020}. For the present study, since the focus is on the influence of $\nu$ and $f$, we operated with the same angle of $\alpha$ = 60$^{\circ}$, with a corresponding tip diameter of $2 r_c$ = 5.8 $\pm$ 0.4 $\mu$m.


The microchannel is fed with liquid seeded with fluorescent and reflective particles (green polystyrene microspheres, Thermo Scientific) of diameter 1 $\mu$m\footnote{The particle diameter has to be much smaller than $\delta$ to get the inner streaming flow, \textcolor{red}{but to measure the amplitude of acoustic vibration velocity and get a qualitative image of the flow (see Fig.~\ref{fig:flow_va}-(b)), larger particles of diameter 4.9 $\mu$m were more adapted.}} by a syringe pump (Newtown Company \& Co). The acoustic wave is ensured by a piezoelectric transducer (Model ABT-455-RC, RS Components) glued on an upper glass microscope coverslip (width $\times$ length $\times$ thickness: 26 mm $\times$ 76 mm $\times$ 1 mm) with epoxy resist. The power is brought by a function generator (Model 33220A, Agilent) with a home-made power amplifier. The transducer spectral response shows several resonance peaks between 400 and 40000 Hz, from which we chose several values of frequency from 500 to 3500 Hz. The applied voltage is sinusoidal, within a range between 0 and 60 V peak-to-peak (up to $\pm$ 30 V). 

The fluids are mixtures of water (W) and glycerin (G) with different rate in W/G. Table \ref{table_liquids} presents the main physical properties of different mixtures used in this study, as well as the values of $\delta$ for the two extreme values of frequency.

\begin{table}
\begin{center}
\begin{tabular}{ccccccc}
\hline\hline
\vspace{0.1 cm}
$w_\mathtt{glyc.}$ & $x_\mathtt{glyc.}$  & $\nu$ (mm$^2$/s) & $c_0$ (m/s) & $\rho_0$ (kg/m$^3$) & $\delta_{3500}$ ($\mu$m)  & $ \delta_{500}$ ($\mu$m) \vspace{0.1 cm} \\ 
\hline\hline
0.00 &  0.00 & 1.007 &  1510 & 998 & 9.57 & 25.3  \\ 
0.062 &  0.05 & 1.158 & 1580 & 1012,7 & 10.3 & 27.1   \\
0.457 & 0.4 &  4.32 & 1760 & 1114.5 & 19.8 & 52.4  \\ 
0.654 & 0.6 &  13.75 & 1810 & 1168.3 & 35.4 & 93.6  \\ 
0.747 &  0.7 & 29.44 & 1840 &  1193.4 & 51.7 & 136.9  \\ 
\end{tabular} 
\end{center}
\caption{Physical properties of water-glycerol mixtures at 20$^{\circ}$C for different mass fraction $w_\mathtt{glyc}$ and volume fraction $x_\mathtt{glyc}$ of glycerol\label{tab: physical parameters}. Data for the viscosity $\nu$ of the water-glycerol mixture are extracted from \cite{Cheng2008}, while the sound speed $c_0$ (at 25$^{\circ}$C) and the density $\rho_0$ are extracted from \cite{Slie1966}. Also indicated are values of the VBL thickness $\delta$ at the highest and lowest frequency $f$, 3500 and 500 Hz.}
\label{table_liquids}
\end{table}

\subsection{Flow visualisation and image processing}

The visualisation is ensured by a fast camera (MotionBLITZ Cube4, Mikrotron) adapted on a binocular microscope. The depth of field of the microscope lens is about 10 $\mu$m, hence five times smaller than the channel depth ($p$ = 50 $\mu$m) which, after careful adjustments, enables to access the streaming velocity near the center plane. A cold-light beam shines from the bottom of the glass slide. While the seeded particles are fluorescent (excitation wavelength 480 nm, light emission wavelength 515 nm), we found that under some conditions of lighting, and due to the limited sensitivity of the camera, the diffused light could offer better contrast than the fluorescence-emitted light. 

By operating under various exposure time and frame-rate from 500 fps to 25000 fps (see details in \cite{Zhang2019}), we can access both the steady streaming velocity $\textbf{V}_s (x,y)$ and the acoustic velocity $\textbf{V}_a (x,y) = \textbf{A} \omega$ (via the vibration amplitude $\textbf{A}$), see Fig.~\ref{fig:flow_va}. In particular, it is observed that close to the tip, $\textbf{V}_s$ can be of the same order as $\textbf{V}_a$. Far from the tip, where the streaming velocity vanishes, the time-cumulated trajectories of individual particles appear as straight segments, along the parallel direction with respect to the channel wall. The measurements of the length of these segments, equal to $2A$, allows to determine the prescribed vibration at infinity $V_a (\infty)$. This appears to us as the most reliable way to quantitatively measure the forcing amplitude, and we denote thereafter for simplicity : $V_a = V_a (\infty)$.
As previously shown \cite{Zhang2019}, the relationship between the prescribed voltage $V$ and the vibration velocity $V_a$ is found to be linear over the range 0 to 60 Volts. For each tested frequency, we proceeded a calibration between voltage and acoustic velocity.

The obtained images are treated with the open-source software ImageJ (\textit{https://imagej.net/}). The streaming velocity field in the plane $(x,y)$ is determined from the relative displacement of particles at a given phase during several vibration periods. Successive frames are converted into displacement vectors and vorticity maps by the software PIVlab (see : \textit{https://pivlab.blogspot.com/}). 

\section{Influence of viscosity}

\subsection{Velocity and vorticity maps}

\begin{figure}
\centering
\subfigure[]{\includegraphics[width=0.48\textwidth]{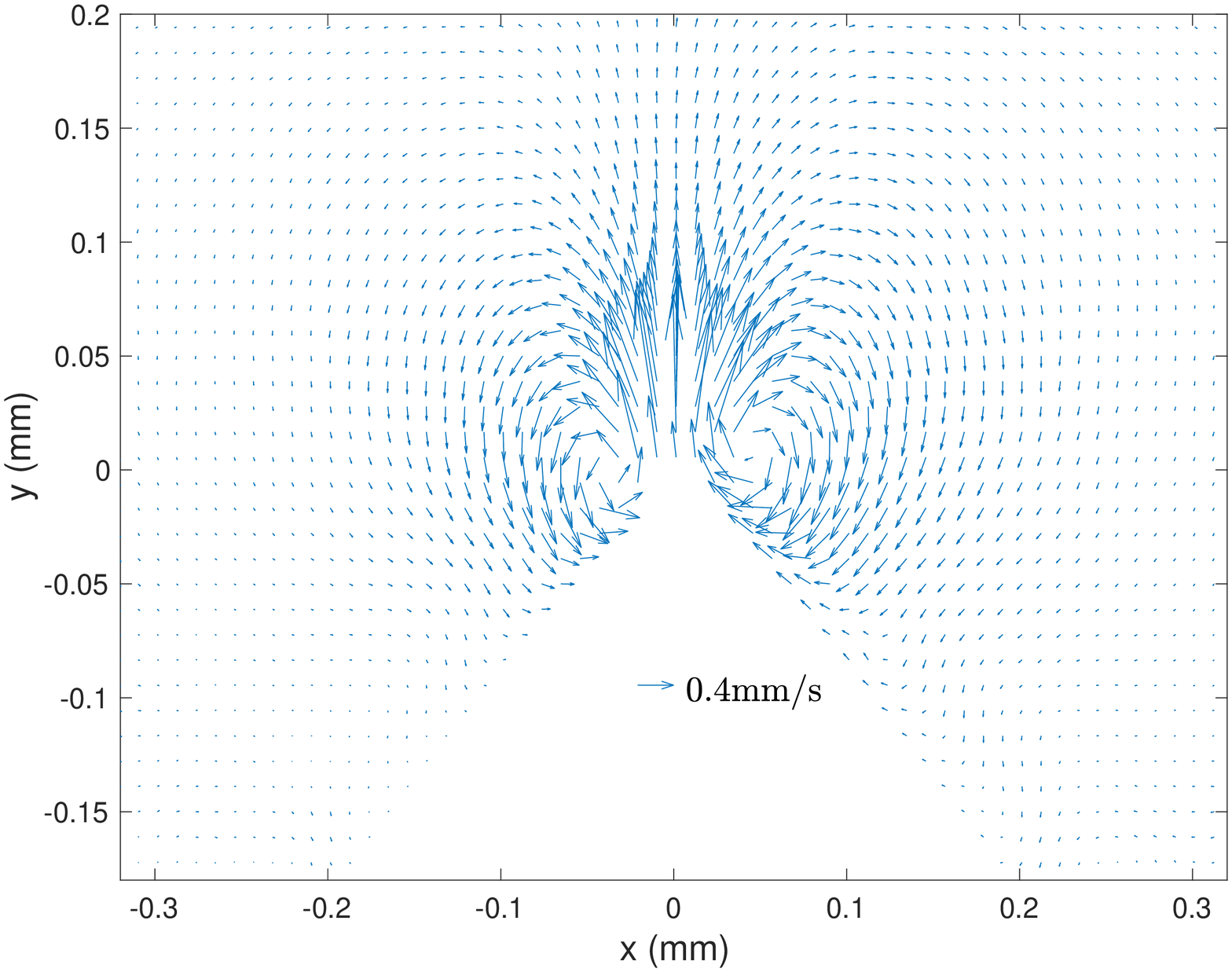}}
\subfigure[]{\includegraphics[width=0.48\textwidth]{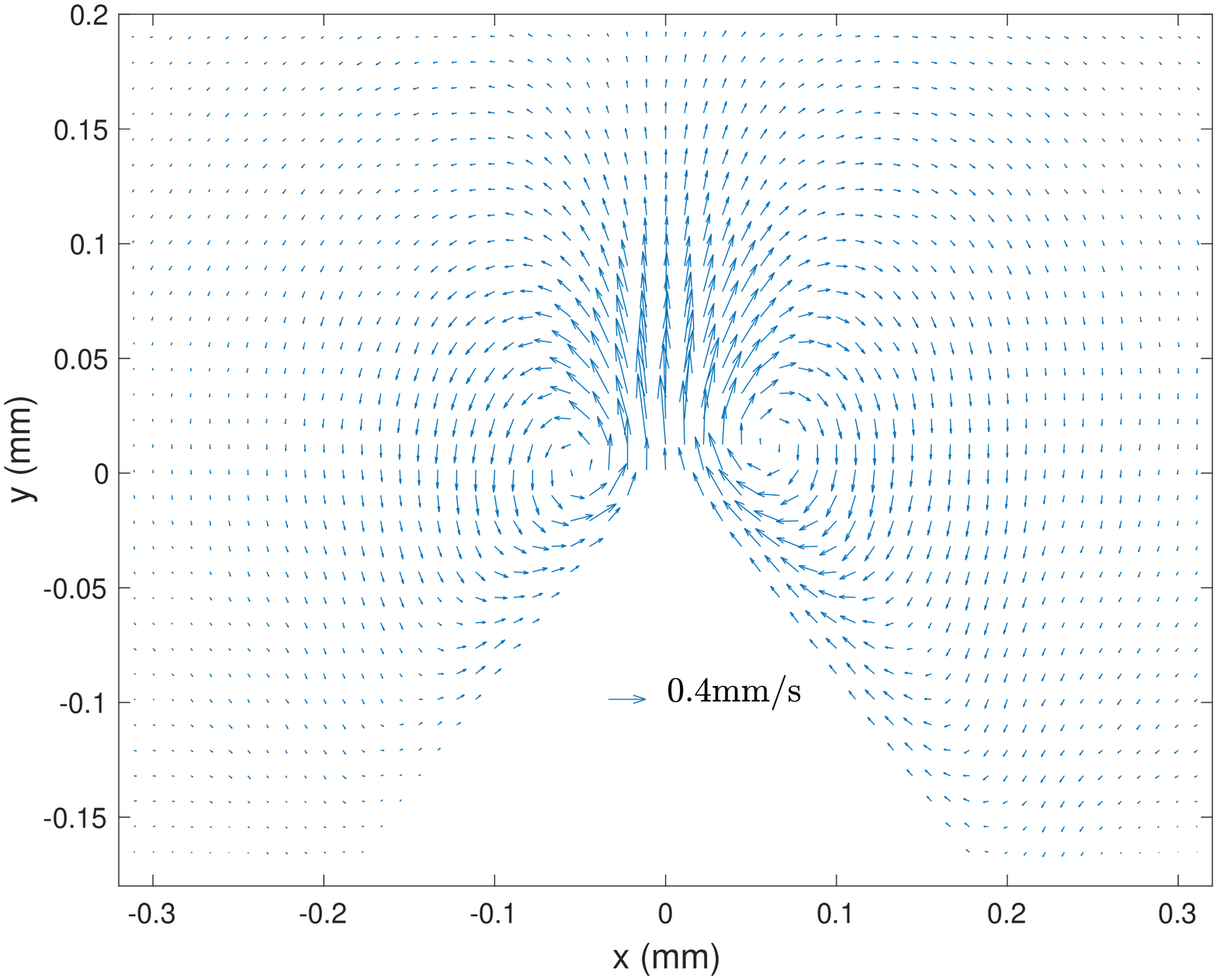}}
\subfigure[]{\includegraphics[width=0.48\textwidth]{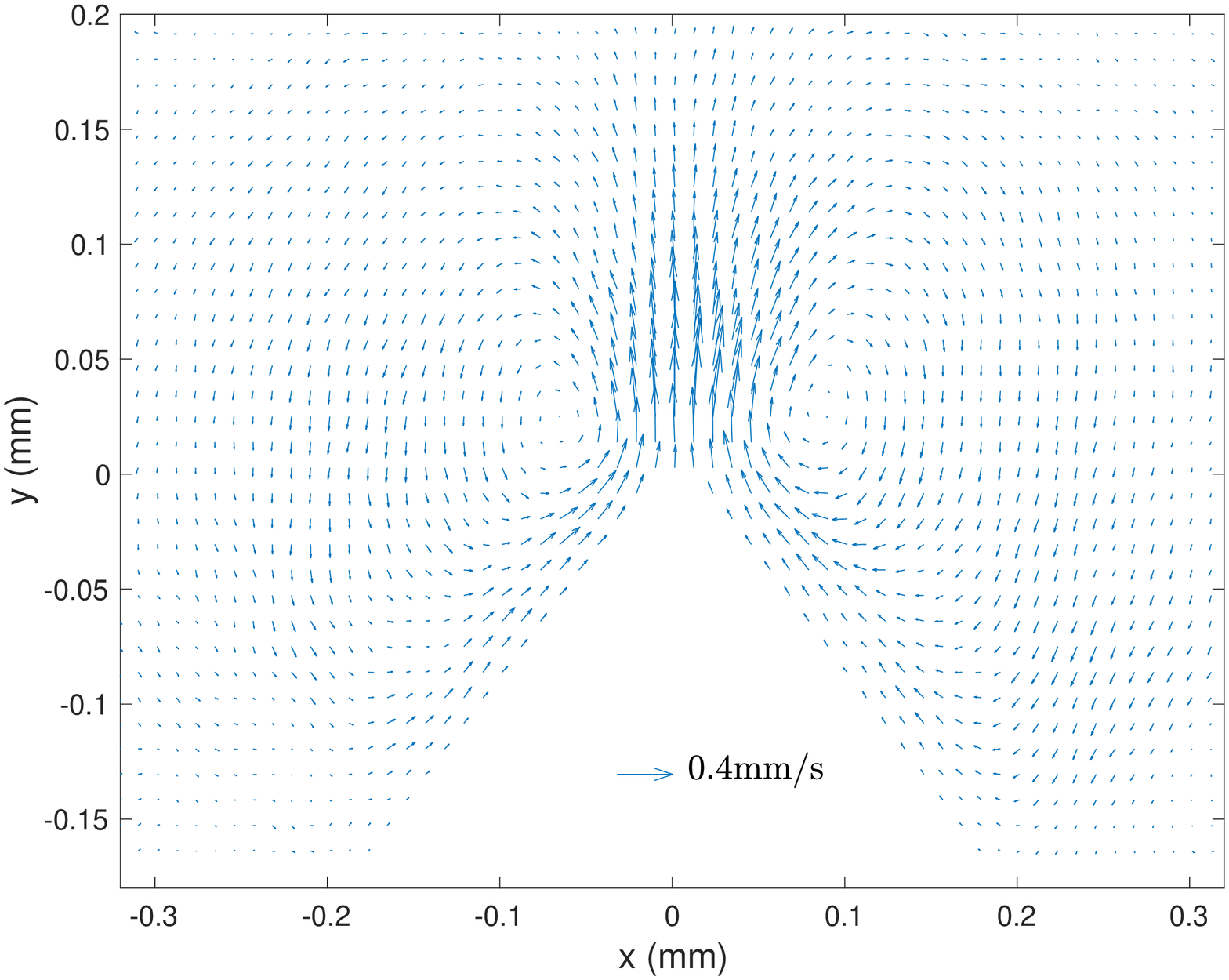}}
\subfigure[]{\includegraphics[width=0.48\textwidth]{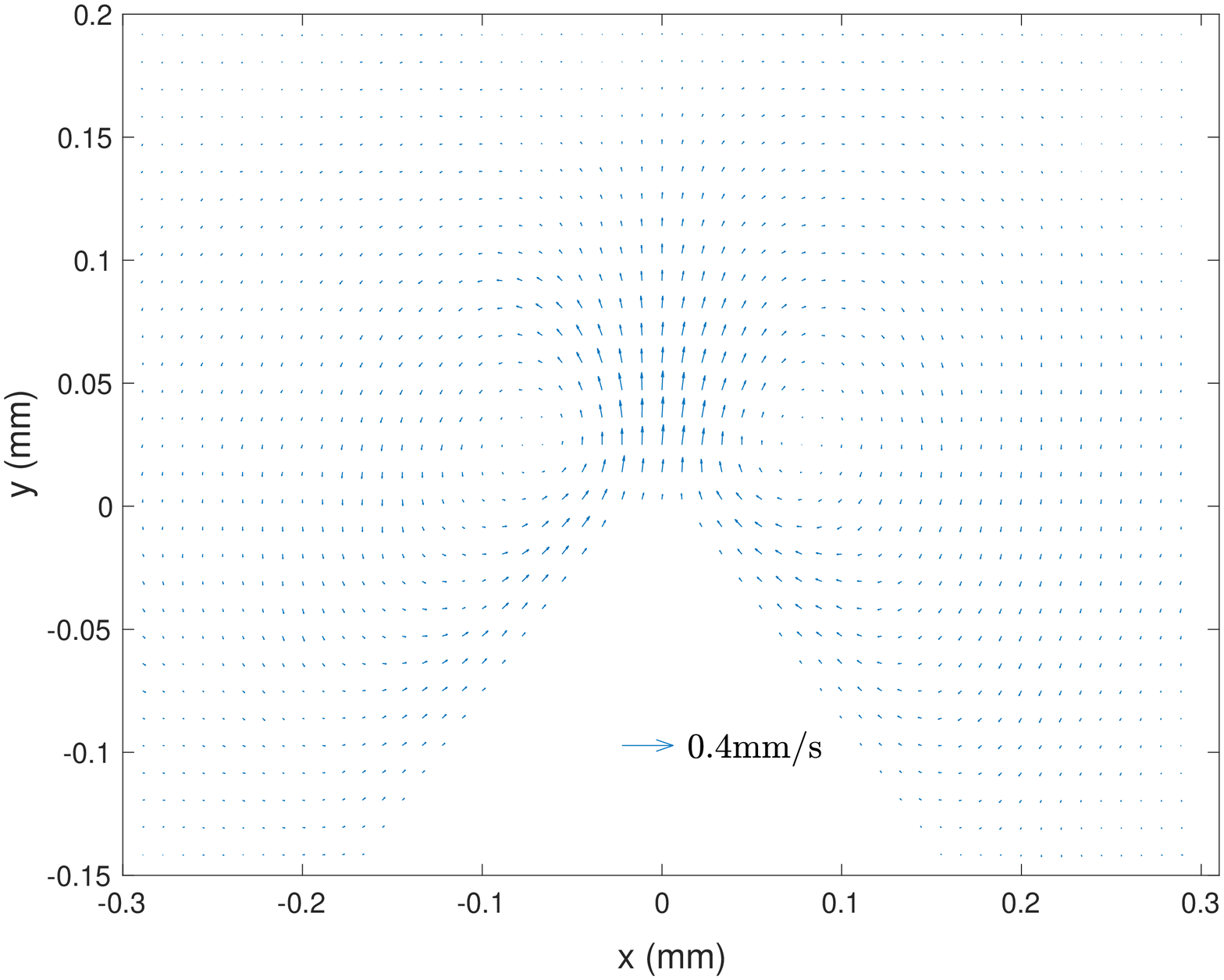}}
\caption{Streaming velocity field $V_s (x,y)$ from PIV measurements, with different liquid viscosities. $f$ = 2500 Hz and $V_a$ = 35 mm/s. (a) $\nu$ = 1.158 mm$^2$/s, (b) $\nu$ = 4.32 mm$^2$/s, (c) $\nu$ = 13.75 mm$^2$/s, (d) $\nu$ = 29.44 mm$^2$/s. Scales are the same for the four cases.}
\label{fig:influ_visco_piv}
\end{figure}

Figures \ref{fig:influ_visco_piv}-(a-d) present typical streaming velocity fields obtained from the PIV treatment. The streaming flow appears as a main central jet from the tip, which is symmetric with respect to the $y$ axis ($x$=0). It clearly appears that the flow intensity decreases with an increasing viscosity. The jet induces the formation of two symmetric vortices beside the sharp edge. In terms of location, the eddies are very near to the tip for the lowest viscosity, and for more viscous liquids they are pushed away and more aside from the tip. Let us also remark that at higher viscosity (figures \ref{fig:influ_visco_piv}-(c and d)), the flow in the VBL along the lateral walls becomes relatively \textcolor{red}{thicker}.

\begin{figure}
\centering
\subfigure[]{\includegraphics[width=0.48\textwidth]{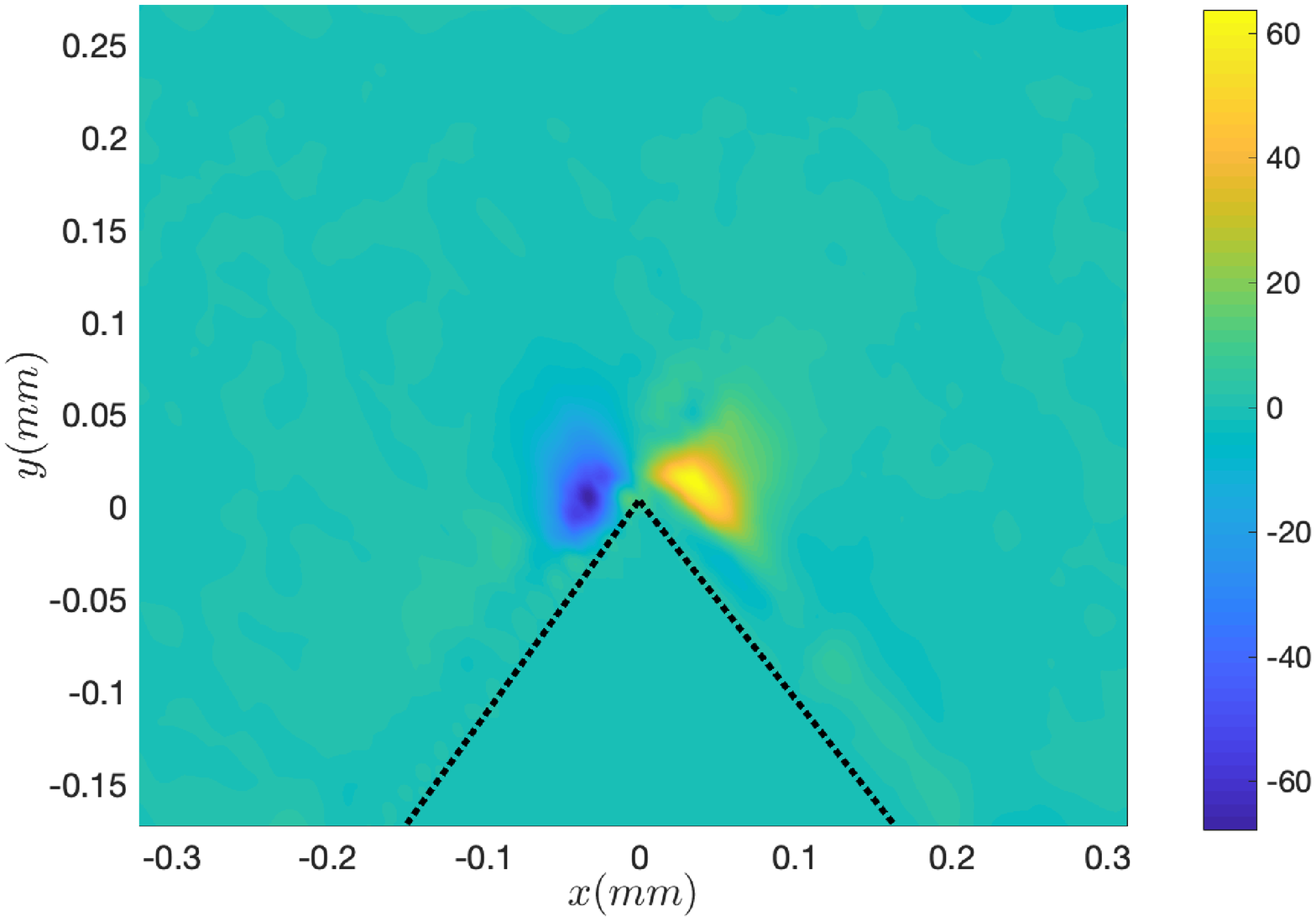}}
\subfigure[]{\includegraphics[width=0.48\textwidth]{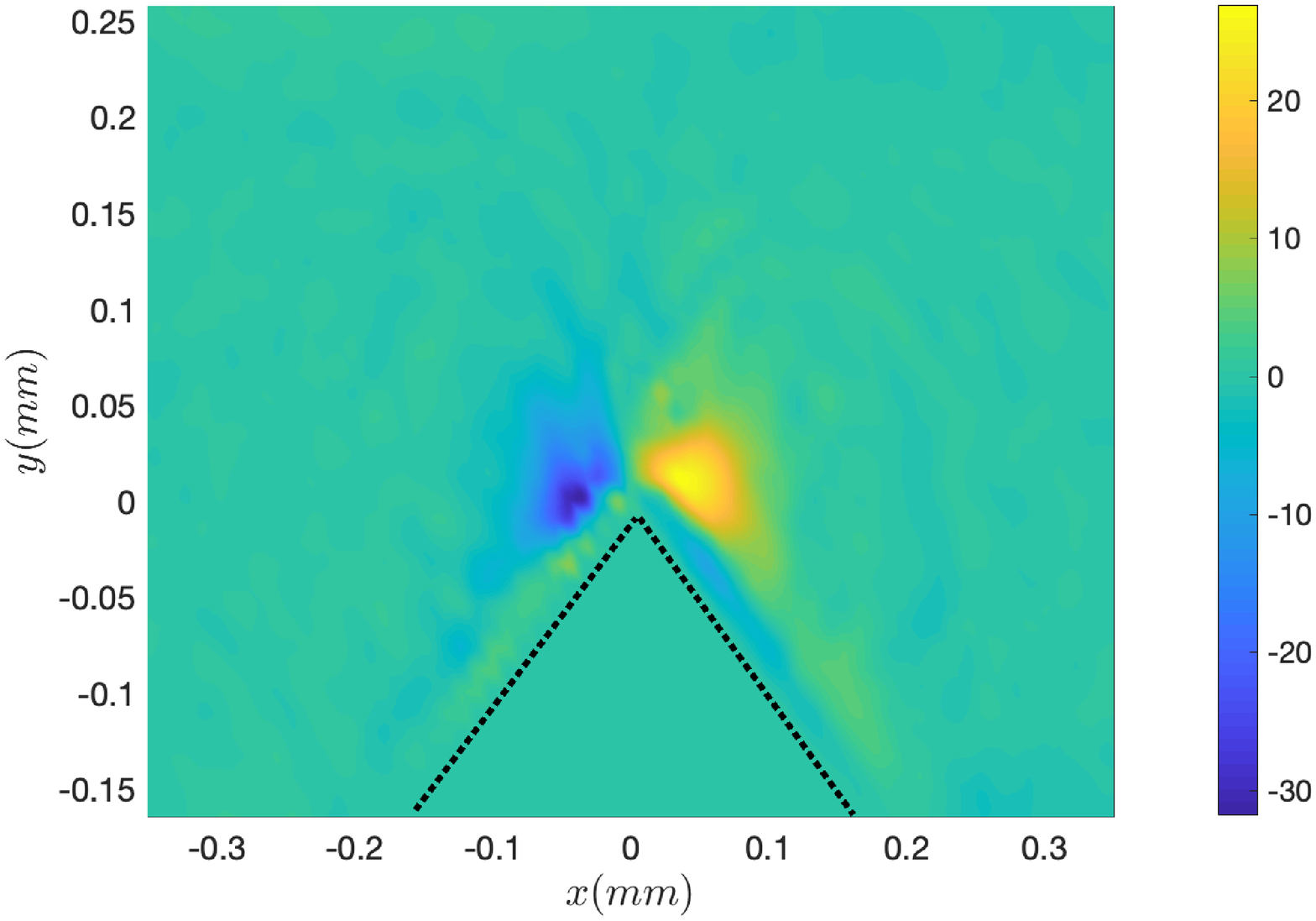}}
\subfigure[]{\includegraphics[width=0.48\textwidth]{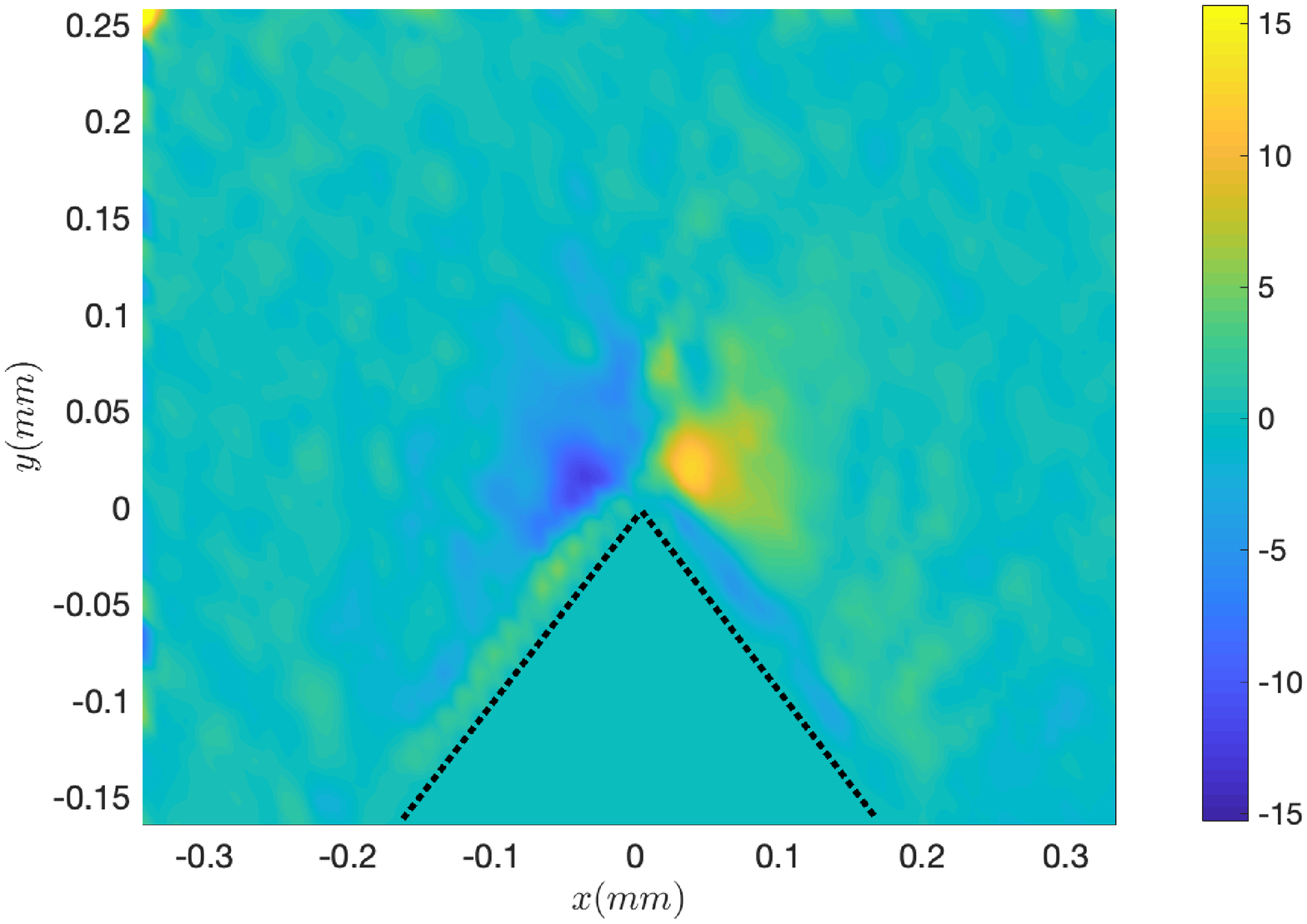}}
\subfigure[]{\includegraphics[width=0.48\textwidth]{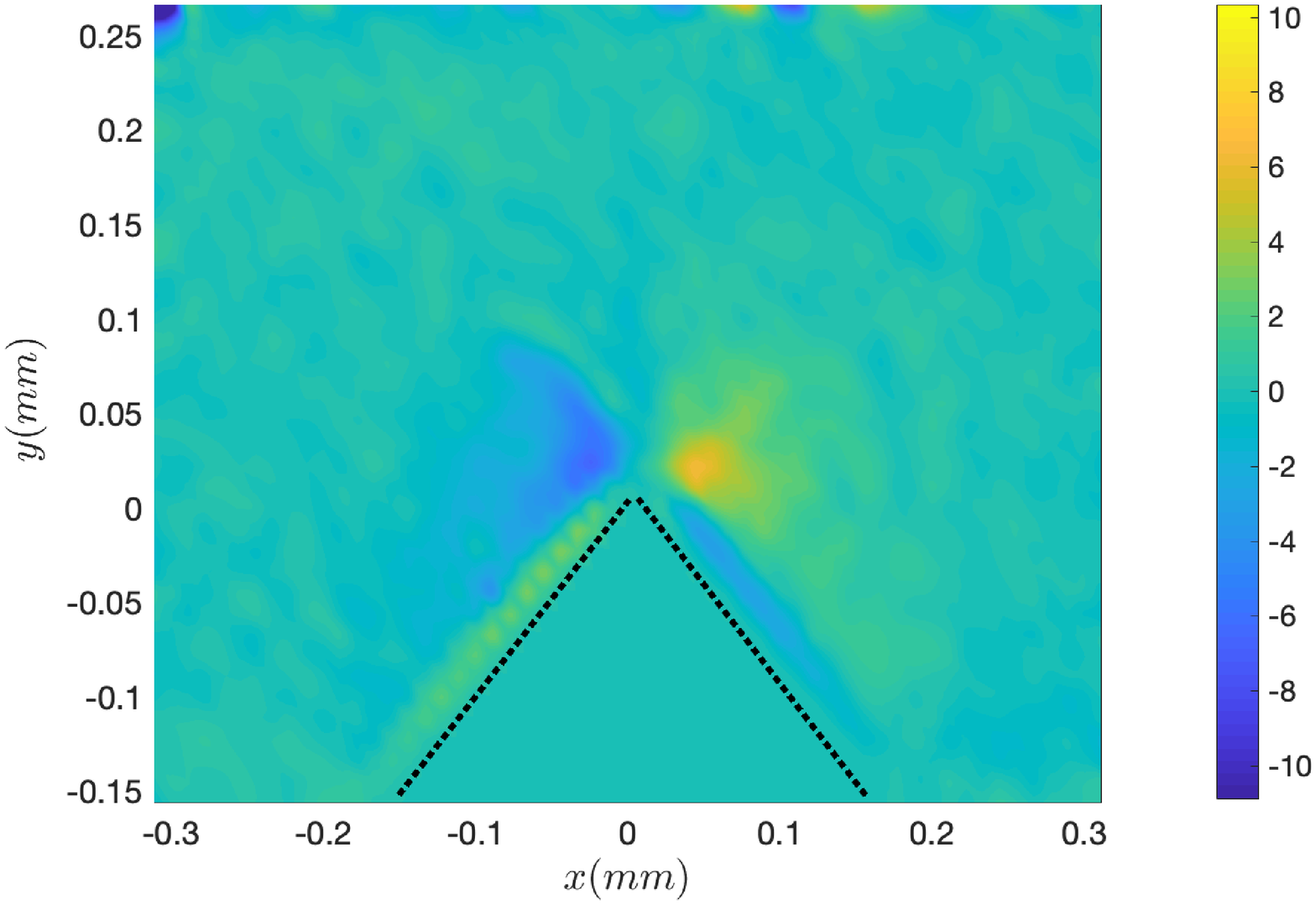}}
\caption{Vorticity maps of the streaming fields corresponding to the cases of figures \ref{fig:influ_visco_piv}-(a-d), with corresponding colorbars that emphasise the decrease of vorticity. Dotted lines show the boundaries of the sharp edge.}
\label{fig:influ_visco_vort}
\end{figure}

Figure \ref{fig:influ_visco_vort} shows the vorticity maps corresponding to the fields of Figures \ref{fig:influ_visco_piv}. The most remarkable point is the decrease of the intensity of the vorticity with increasing viscosity, as testified by the scales of the colormaps from (a) to (d). However, the size of the vortices, which may characterise the disturbance distance, remains roughly equal for all liquid samples. Also, the thickness of the inner vorticity areas, and the absolute vorticity within this specific region, appear to be roughly constant for all liquids. 

\subsection{Maximal streaming velocity at different viscosities}

\begin{figure}
\centering
\includegraphics[width=0.6\textwidth]{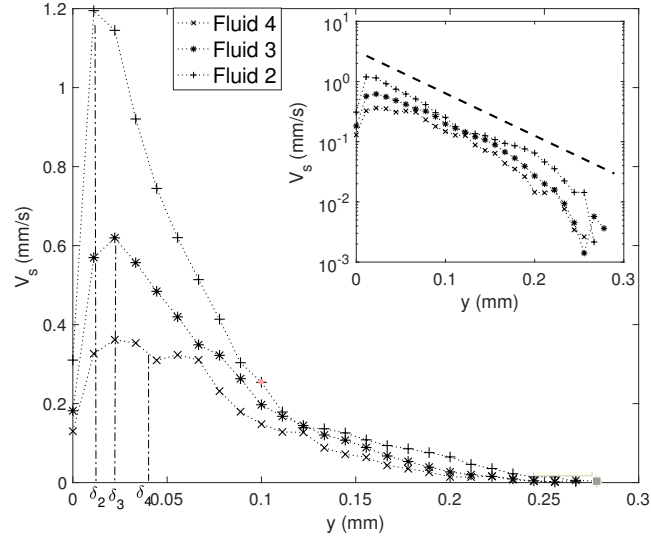}
\caption{Streaming velocity profile along vertical direction $V_s (y)$, for three different viscosities (Fluids 2, 3 and 4 with $\nu$ respectively equal to 1.158, 4.32 and 13.75 mm$^2$/s.). The operation condition is at frequency $f$= 2500 Hz and acoustic velocity $V_a$ = 35 mm/s. Also labelled, the values of the VBL thickness for the three fluids $\delta_2$, $\delta_3$ and $\delta_4$. The inset plots the same data in Lin-Log axes.}
\label{fig:influ_visco_3fluids}
\end{figure}

To further quantify the flow pattern, we extract the flow profile along the $y$ axis: $V_s (x=0,y)$, for different viscosities and forcing amplitudes. Figure \ref{fig:influ_visco_3fluids} shows three examples of profiles for the same $V_{ a}$ = 35 mm/s and fluids 2, 3 and 4 (see Table 2). It shows a quantitative confirmation that a higher viscosity entrains less intense and relatively more spread profiles. Since the velocity fields are symmetrical with respect to the $y$ axis, the maximal velocity $V_{s max}$ can be extracted from these profiles. It turns out that the maximal velocity is roughly located at a distance $y= \delta$ from the tip.

A more careful examination of the decaying of $V_s (x=0, y)$ suggests that the influence of viscosity is mainly significant within a region of a few VBL thickness. Conversely, the decaying zone further from the tip seems to follow a decreasing exponential behavior, which is almost independent of $\nu$ : the profiles are just shifted from each other by a velocity offset. In addition, at a distance of roughly 130 $\mu$m, $V_s (x=0, y)$ approaches zero for all cases. This length scale seems to depends only on the sharp edge structure, which is in our case characterised by an angle of 60$^{\circ}$, and tip height $h$= 180 $\mu$m. 

\begin{figure}
\centering
\includegraphics[width=0.48\textwidth]{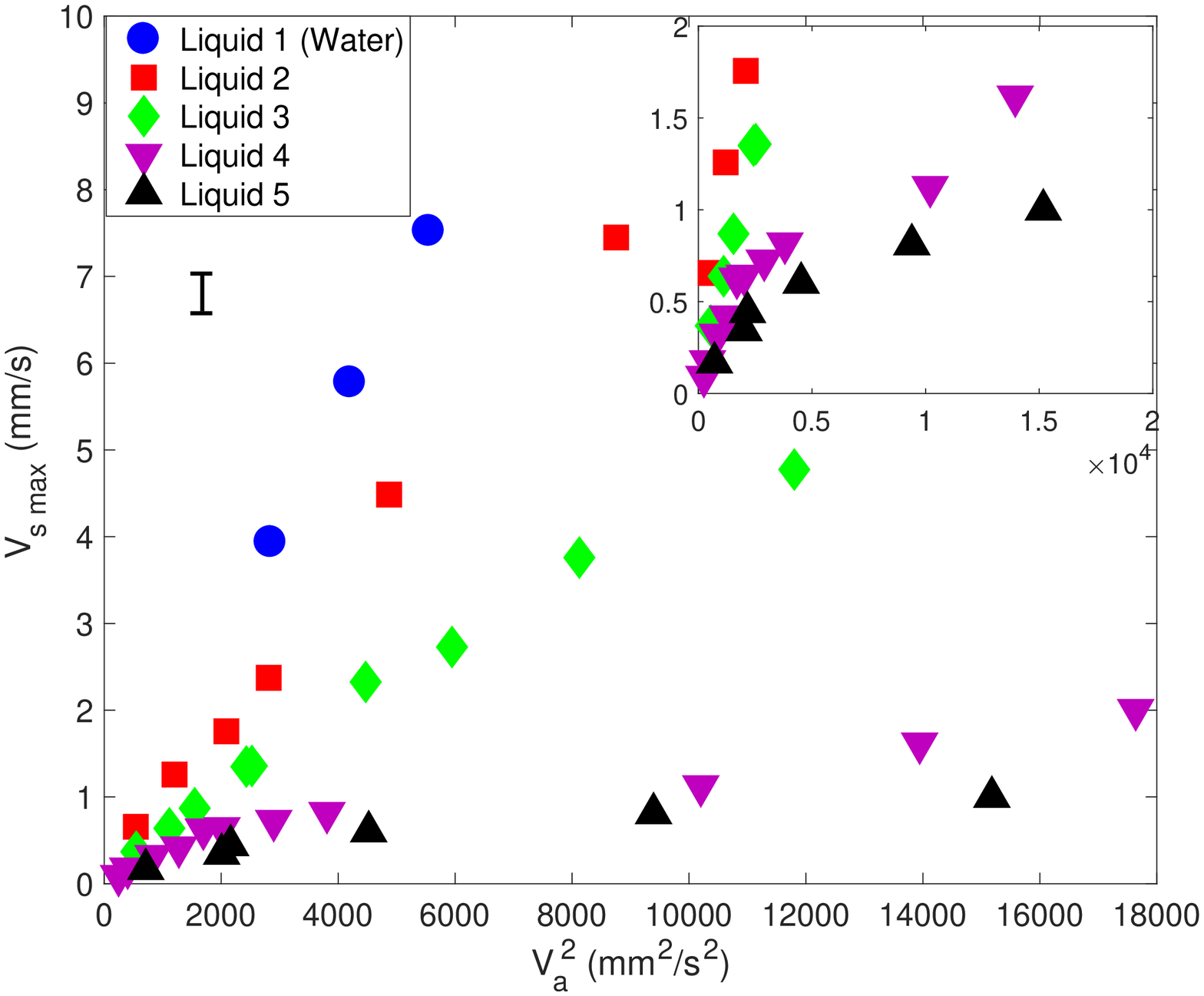}
\includegraphics[width=0.48\textwidth]{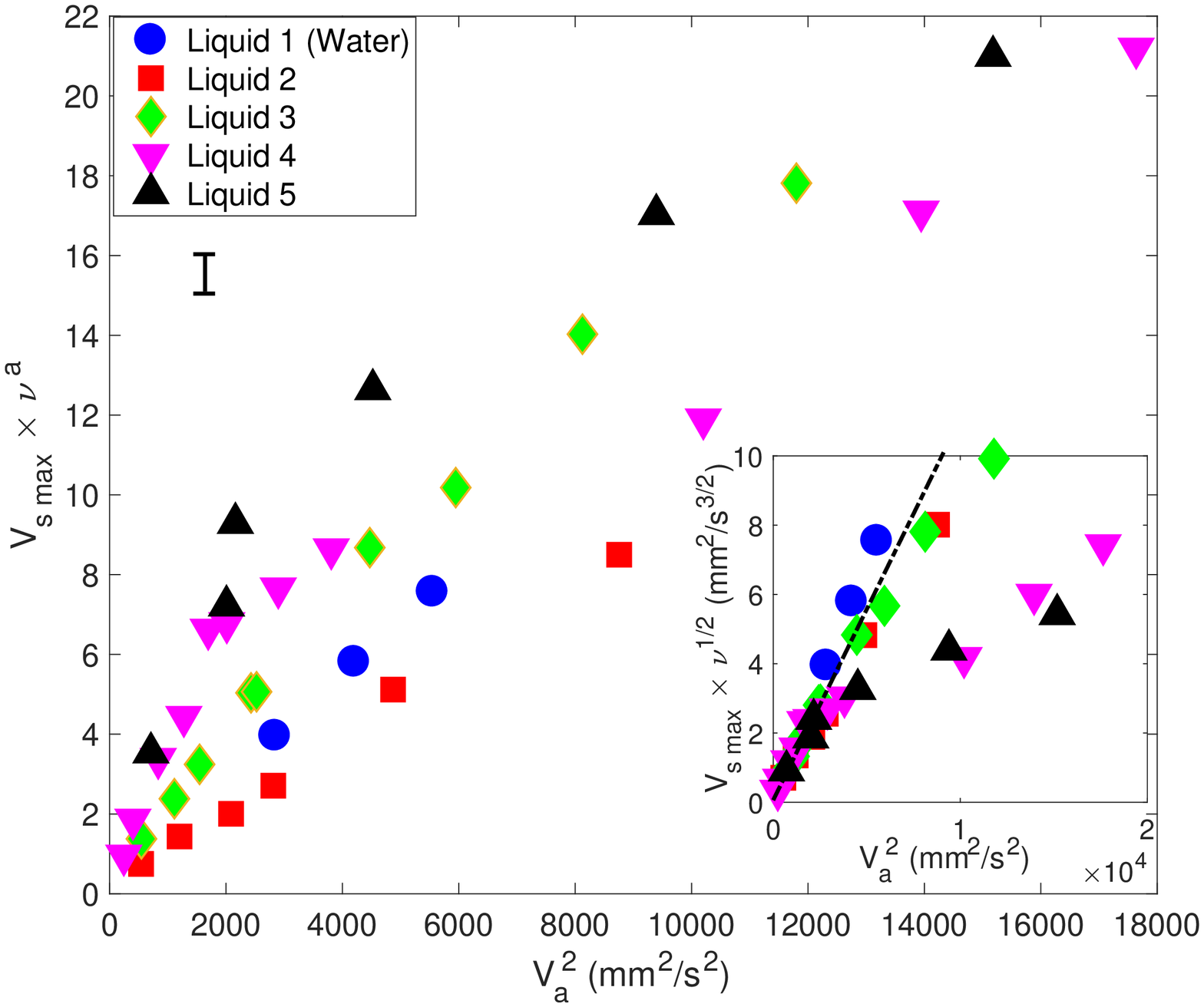}
\caption{\textit{Left} - Maximal streaming velocity $V_\text{{s max}}$ versus the square of the acoustic forcing velocity $V_a^2$, for different liquid viscosities $\nu$, indicated in Table 2. \textit{Right} - Quantity $V_\text{{s max}} \times \nu^{-a}$, with $a$= - 0.9. Inset $V_\text{{s max}} \times \nu^{1/2}$. All measurements were obtained at $f$ = 2500 Hz. The averaged typical error bar is indicated.}
\label{fig:influ_visco}
\end{figure}  

Now we focus on the measurements obtained within a large range of $V_a$. Quantitatively, we mainly focus on the maximal - and characteristic, value of $V_s(x,y)$ measured around $y=\delta$ and at $x$ = 0. In what follows, we shall also extract the prefactor $\theta$ that relates $V_s$ to $V_a^2$, from the whole data set where the dependence is linear. Back to eq.~(\ref{eq:ovchinnikov_vs}), $\theta$ is equal to $\frac{1}{\nu}\frac{\delta^{2n-1}}{a^{2n-2}}$, from which the dependence on $\nu$ and on $f$ can be readily predicted, taking $\alpha$ = 60$^{\circ}$ as in our experiments : 

\begin{equation}
V_s \sim  \nu^{-0.9} f^{-0.1} ~ V_a^{ 2}
\label{eq:th_predict}
\end{equation}

To verify this theory, figures \ref{fig:influ_visco} show the results of the experimental maximal streaming velocity $V_\text{{s max}}$ versus the square of the acoustic forcing velocity amplitude $V_a^{ 2}$, presented either as raw data (Left) or via the quantity $V_\text{{s max}} \times \nu^{-a}$, with $a$ is an exponent deduced from Ovchinnikov \textit{et al.}'s theory \cite{Ovchinnikov2014}, equal to -0.9 for an angle $\alpha$ = 60$^{\circ}$ as stated above. In the inset, the quantity $V_\text{{s max}} \times \nu^{1/2}$ plotted versus $V_a^{ 2}$, shows a partial collapse of data in the range of the smallest values of $V_a^{ 2}$, roughly below 800 mm$^2$/s. At this stage of our investigations, we are unable to explain such a trend. From these results, we can simply conclude that viscosity strongly influences the streaming flow generated around sharp edges. But the dependence cannot be simply captured by the predictions of the perturbative theory from Ovchinnikov \textit{et al.} \cite{Ovchinnikov2014}, nor by any arbitrary power-law. In any case, the results show the quantitative confirmation that the independence on $\nu$ observed in classical Rayleigh-Schlichting streaming is lost in sharp-edge streaming.

Let us finally point out that for more viscous liquids (4 and 5), there is a clear departure from a linear dependence between $V_{s max}$ and $V_a^{ 2}$, typically as $V_a^{ 2}$ is larger than roughly 800 mm$^2$/s. For these two liquids, at 2500 Hz, $\delta_4$ = 41.8 $\mu$m and $\delta_5$ = 61.2 $\mu$m, hence that $p^*$ is of the order of one.

\section{Influence of frequency}

\subsection{Velocity and vorticity maps}

\begin{figure}
\centering
\subfigure[]{\includegraphics[width=0.48\textwidth]{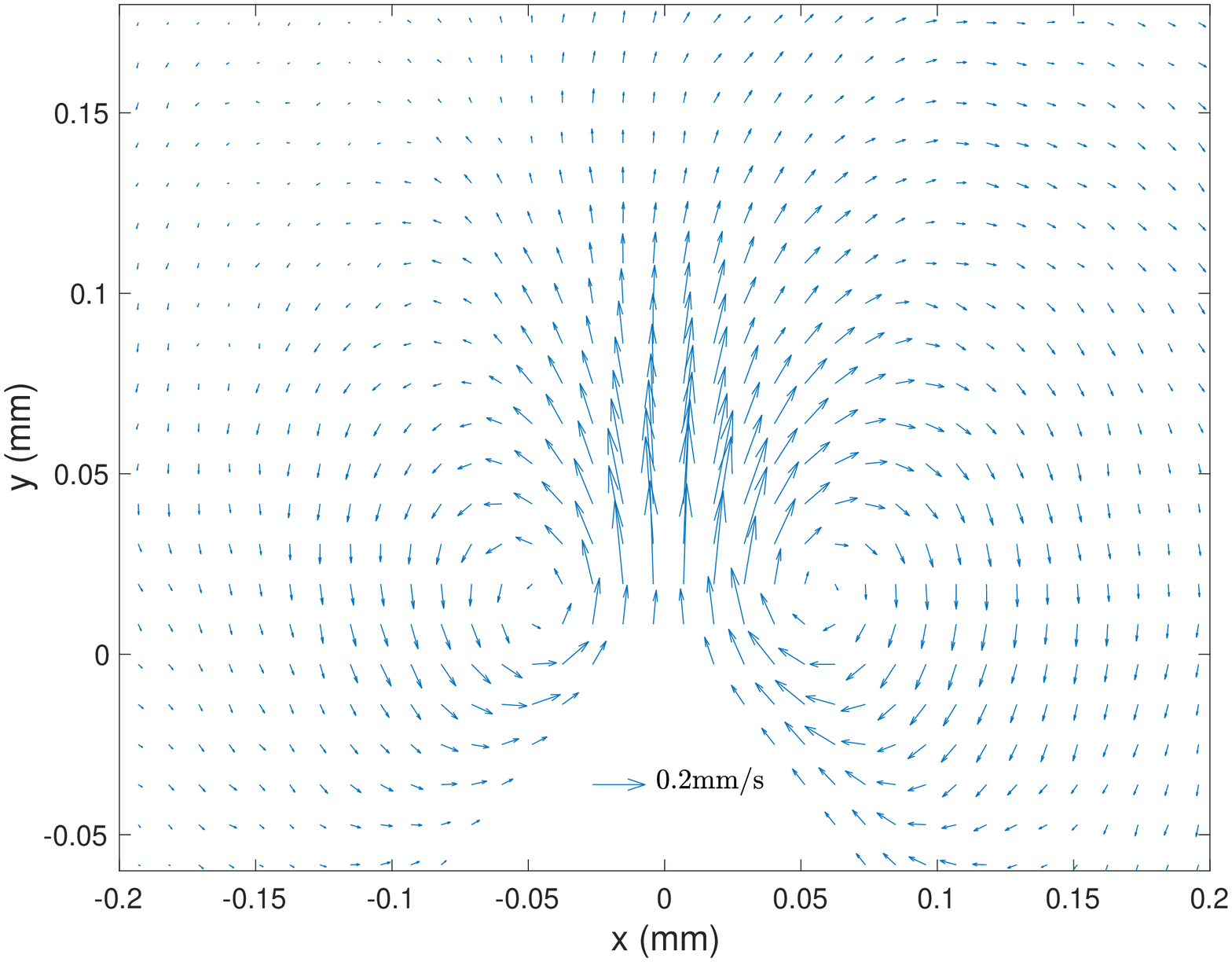}}
\subfigure[]{\includegraphics[width=0.48\textwidth]{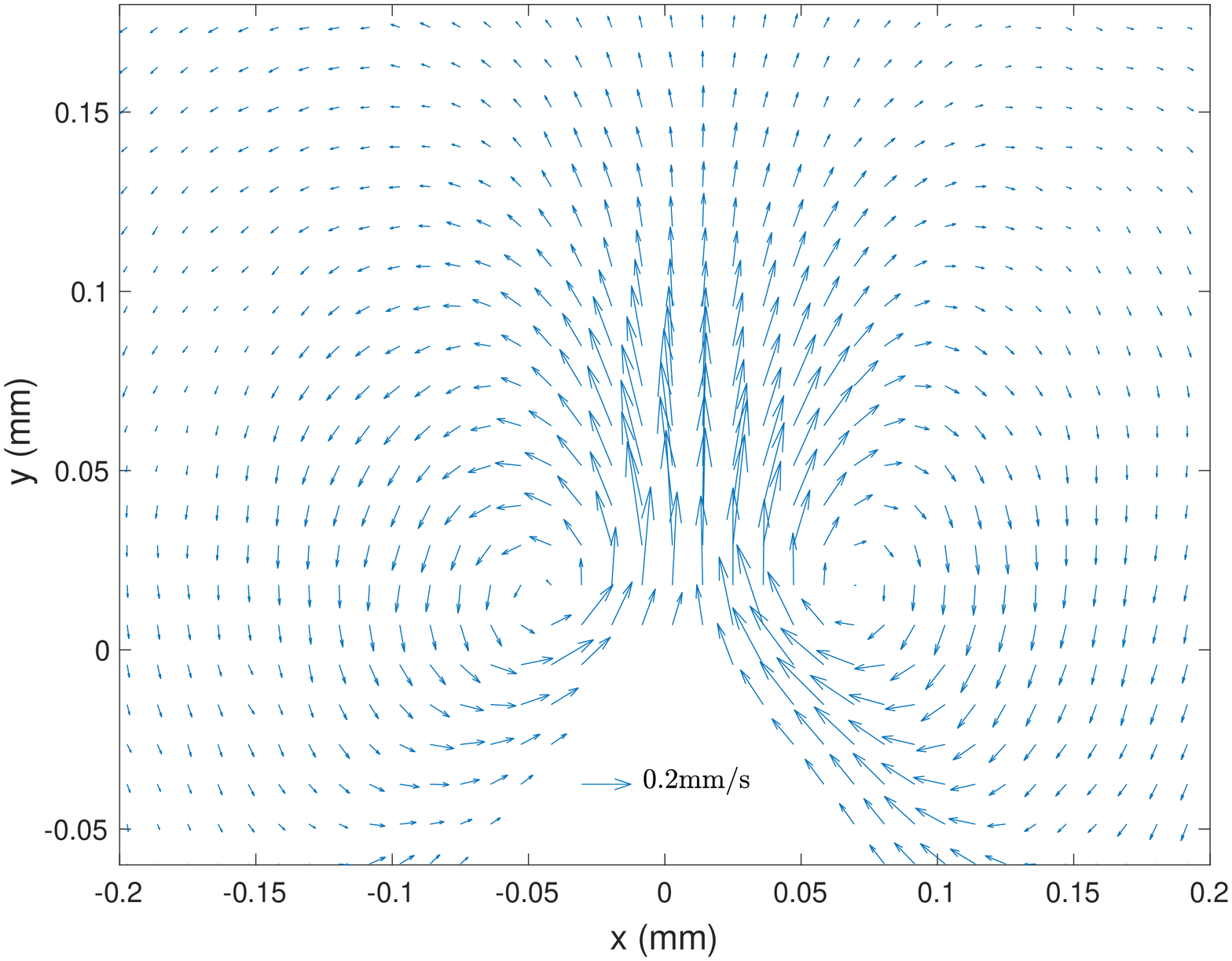}}
\subfigure[]{\includegraphics[width=0.48\textwidth]{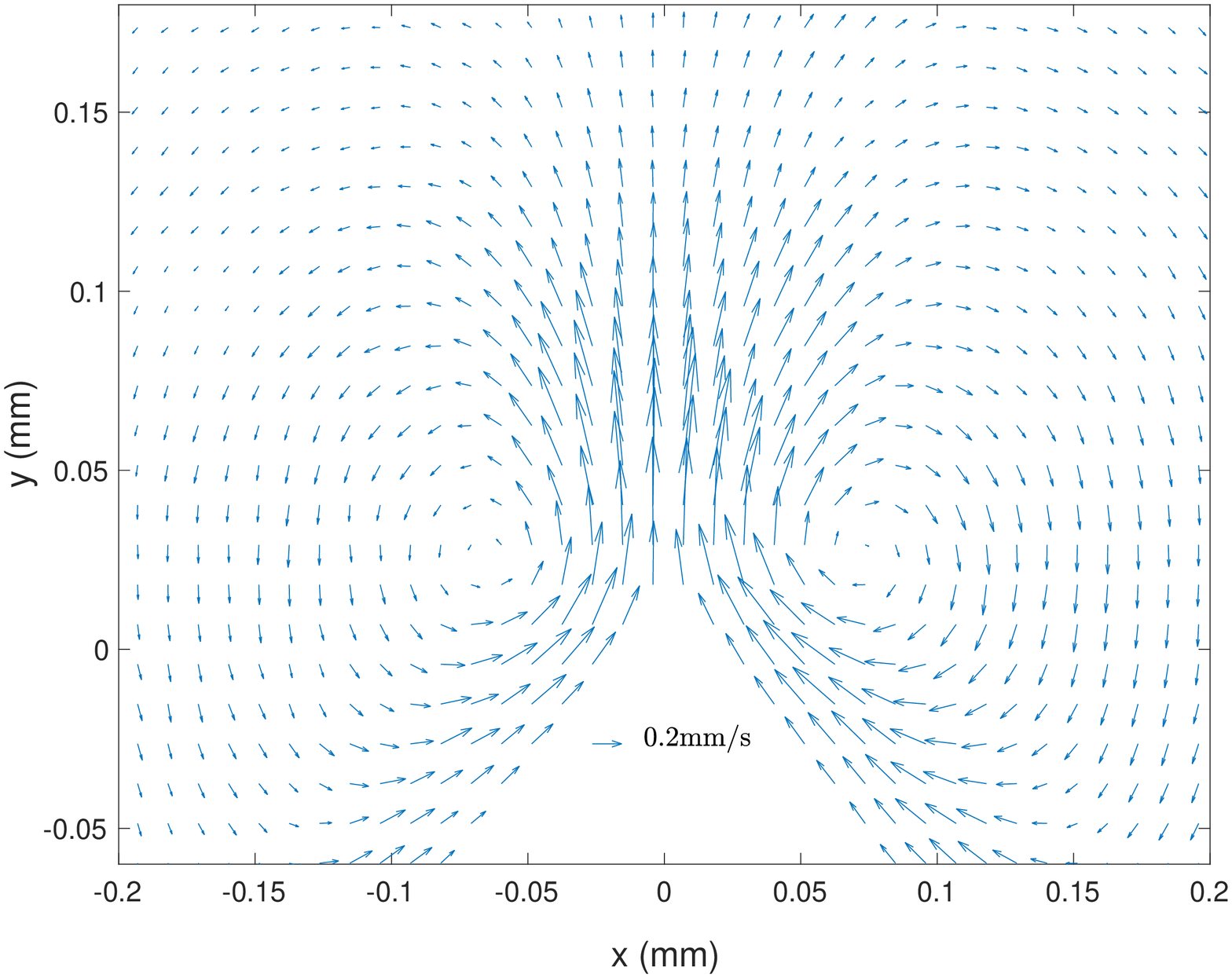}}
\subfigure[]{\includegraphics[width=0.48\textwidth]{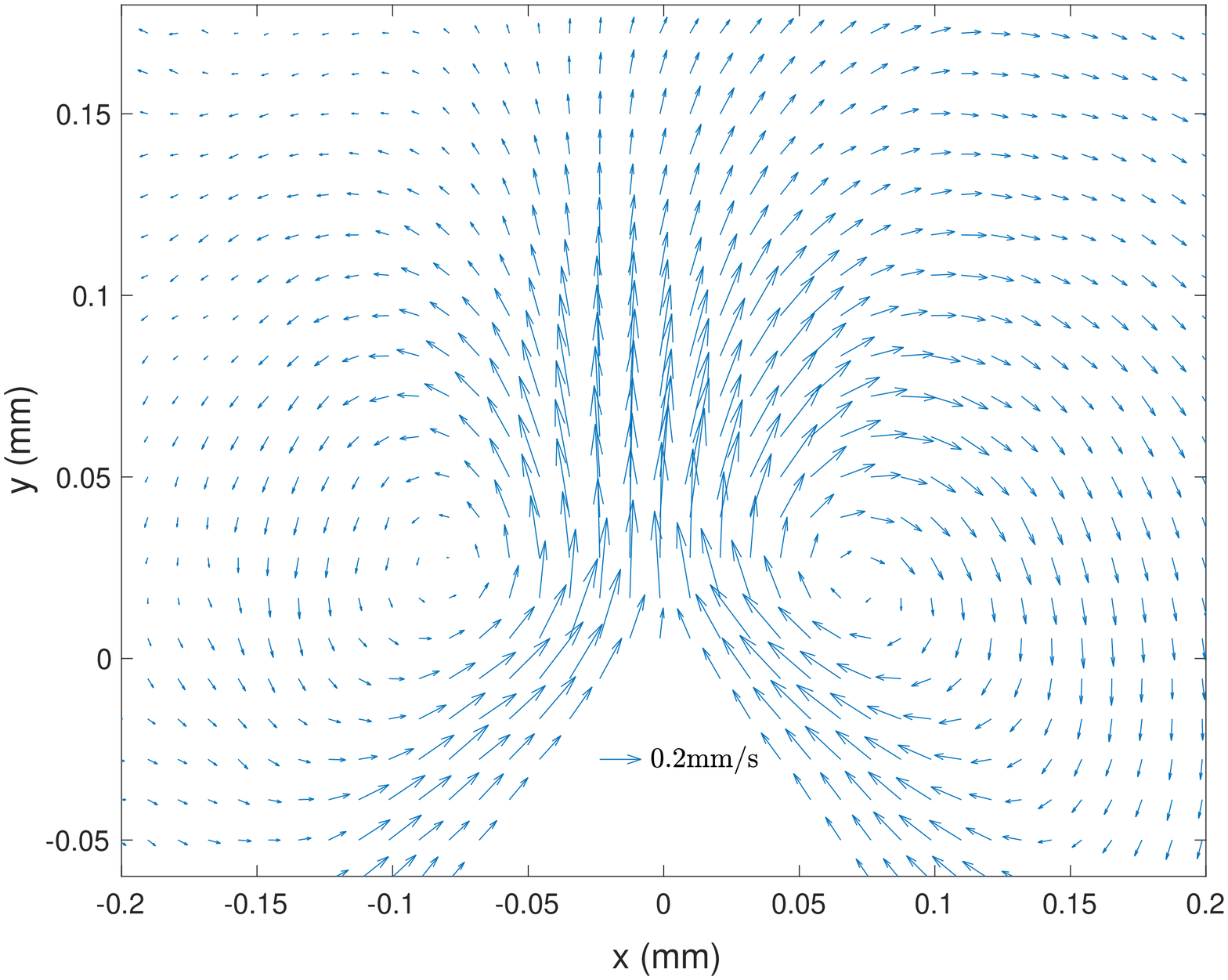}}
\caption{Streaming velocity field $V_s (x,y)$ from PIV measurements, with different excitation frequencies $\nu$ = 4.32 mm$^2$/s (Fluid 3) and $V_a$ = 22.4 mm/s. (a) $f$ = 3500 Hz, (b) $f$ = 2500 Hz, (c) $f$ = 1250 Hz, (d) $f$ = 800 Hz. Scales are the same for the four cases.}
\label{fig:influ_freq_piv}
\end{figure}

\begin{figure}
\centering
\subfigure[]{\includegraphics[width=0.48\textwidth]{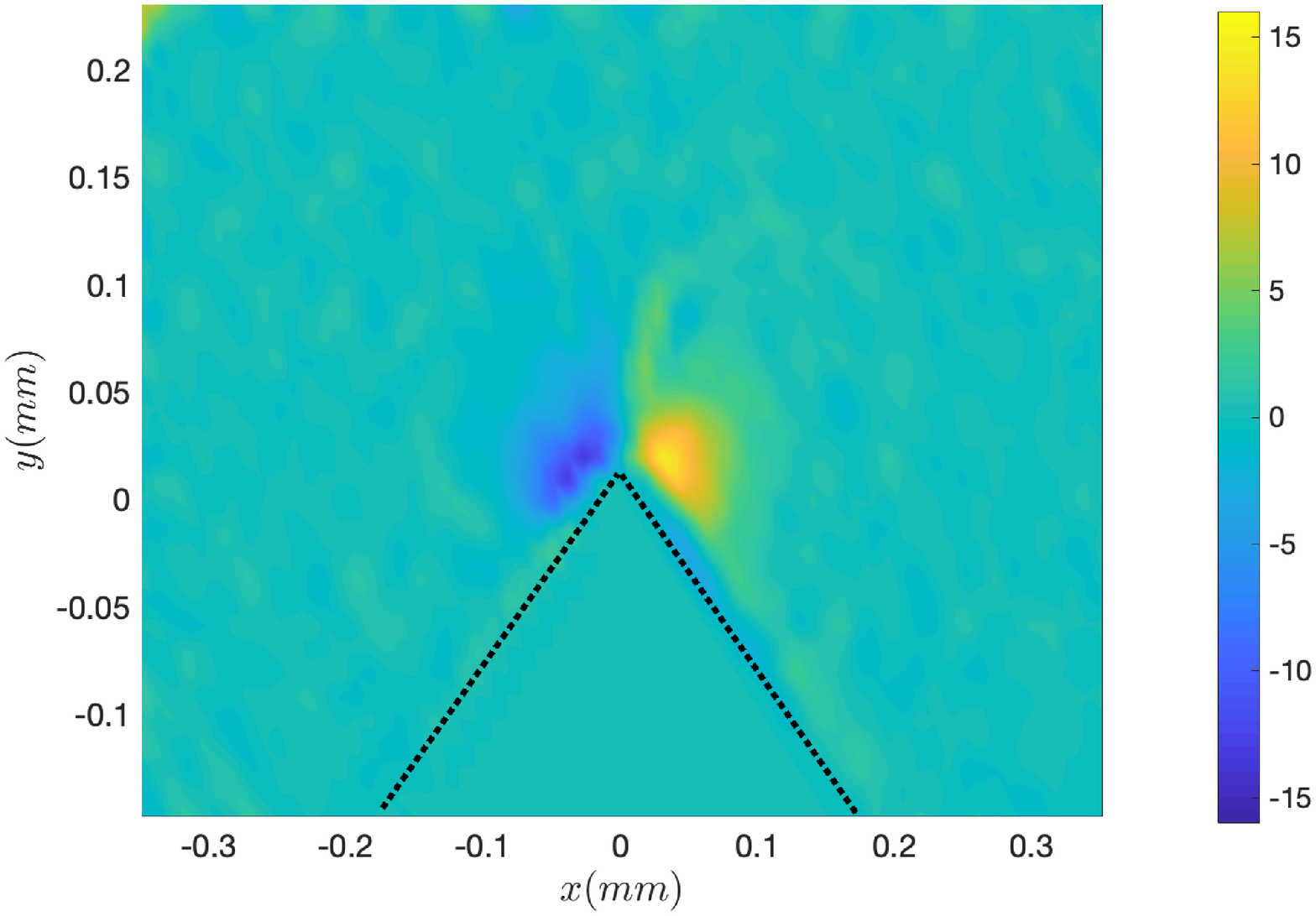}}
\subfigure[]{\includegraphics[width=0.48\textwidth]{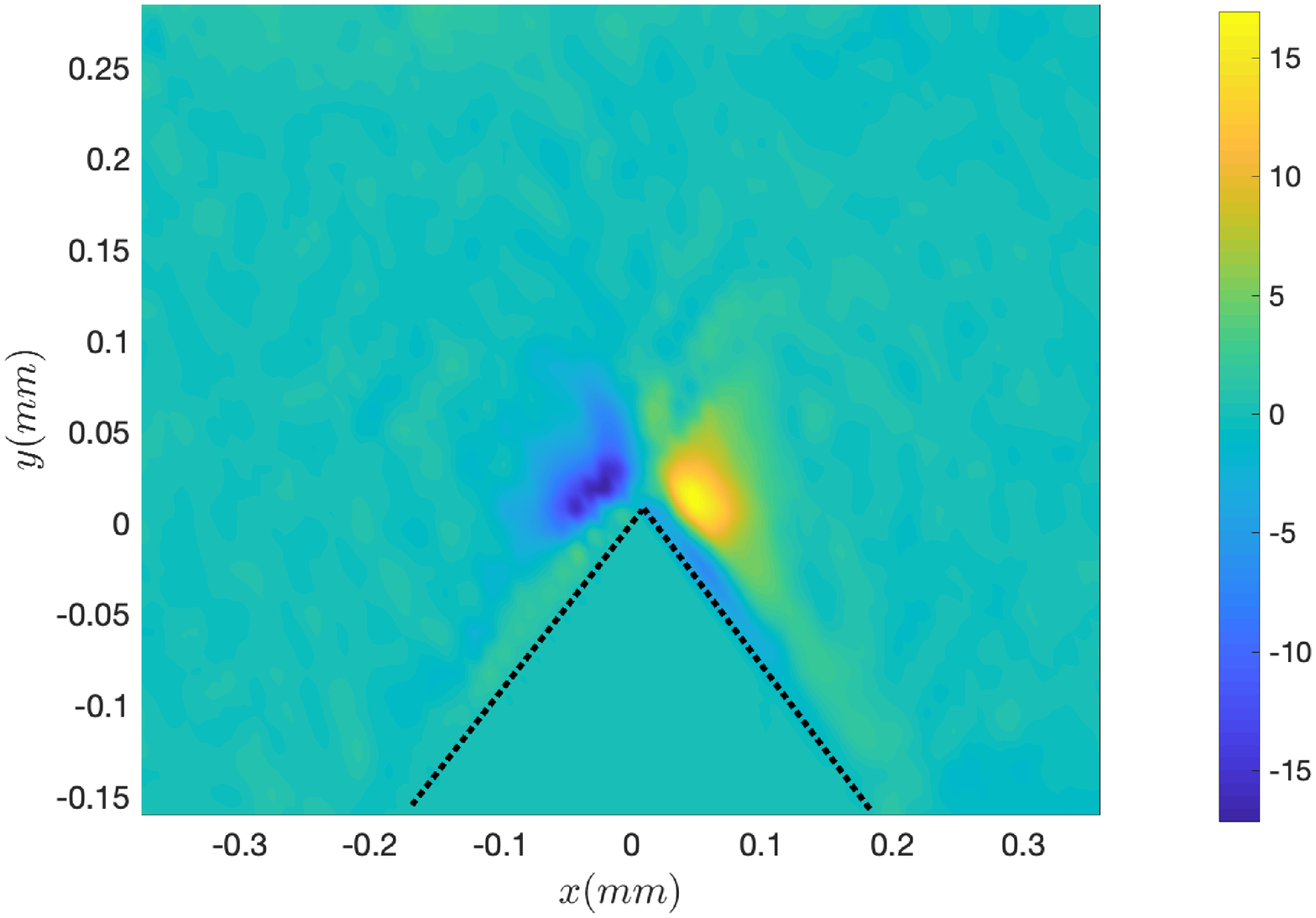}}
\subfigure[]{\includegraphics[width=0.48\textwidth]{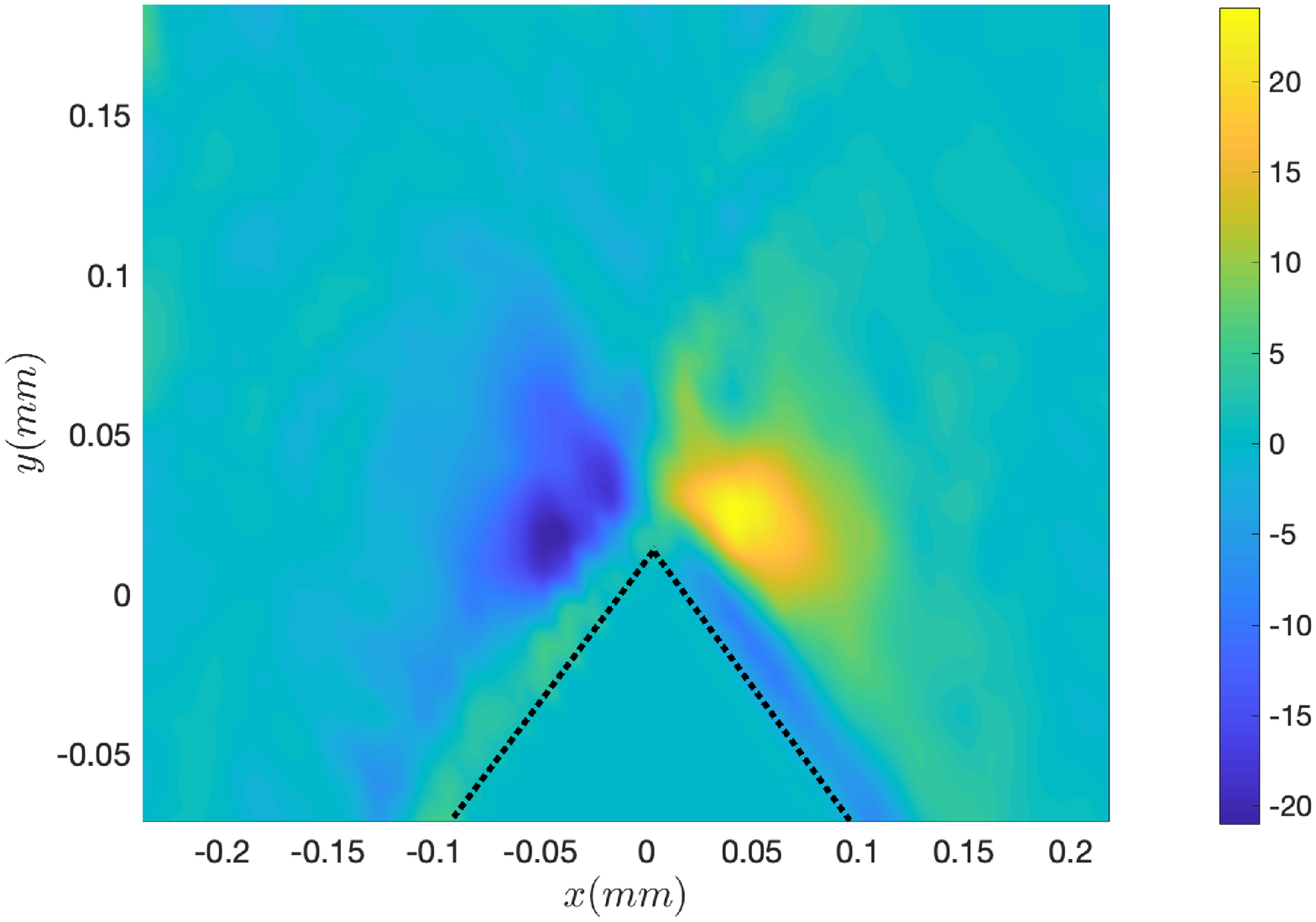}}
\subfigure[]{\includegraphics[width=0.48\textwidth]{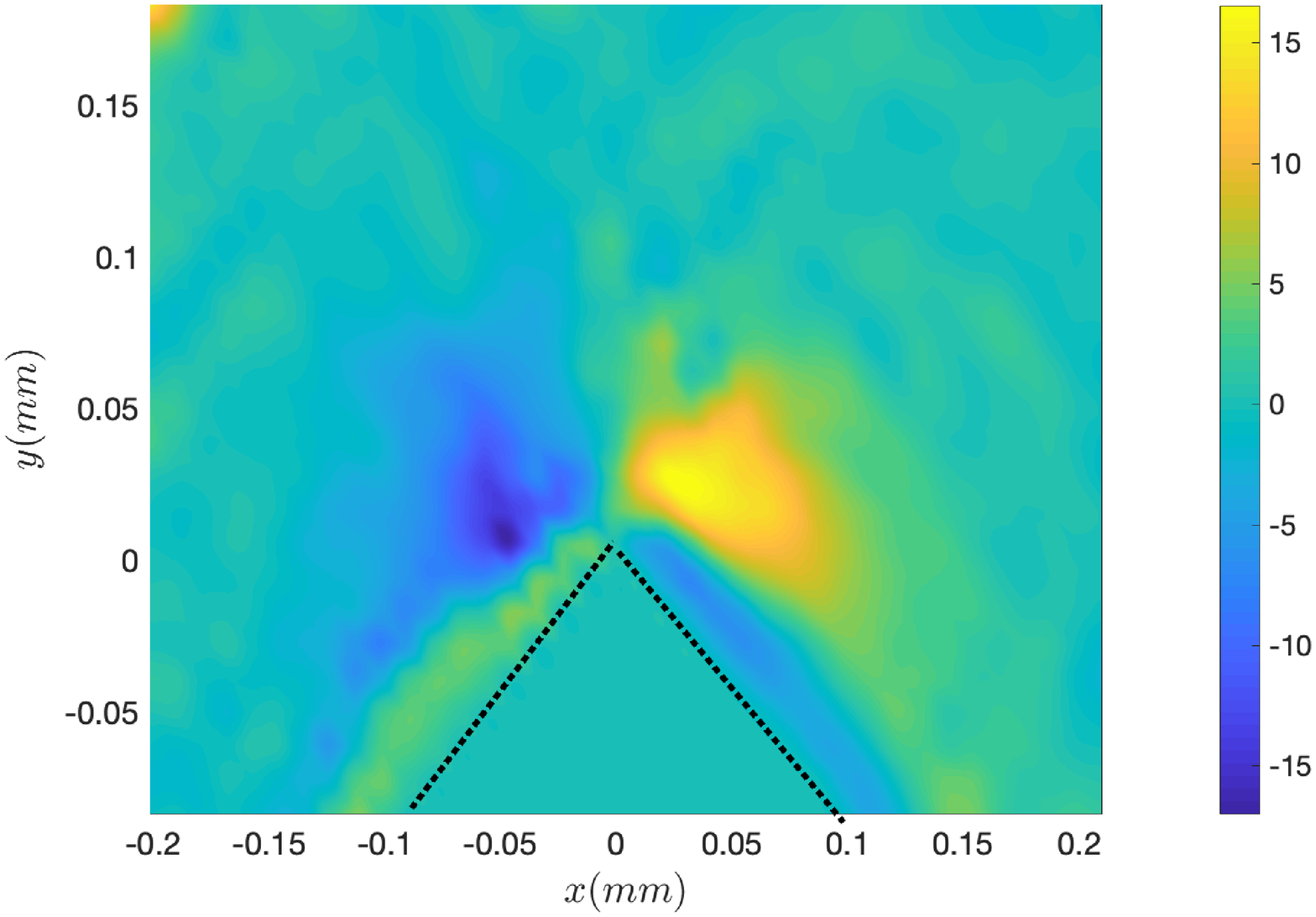}}
\caption{Vorticity maps of the streaming fields corresponding to the cases of figures \ref{fig:influ_freq_piv}-(a-d). Dotted lines show the boundaries of the sharp edge.}
\label{fig:influ_freq_vort}
\end{figure}

Figures \ref{fig:influ_freq_piv}-(a-d) present typical streaming velocity fields at different frequencies ($f$ = 3500, 2500, 1250 and 800 Hz) with the same liquid viscosity ($\nu$ = 4.32 mm$^2$/s) and forcing amplitude ($V_a$ = 22.4 mm/s). The same global structure with the main central jet and the inner and outer vortices are observed for all frequencies. Though, the frequency does not seem to influence significantly the order of magnitude of the flow. Figures \ref{fig:influ_freq_vort}-(a-d) show the corresponding vorticity maps. Let us note that the colormap scale is comparable for all four frequencies. As frequency gets lower, one observes a thicker and more intense inner VBL along the walls, while the outer vortices are more spread. The magnitude of vorticity in the outer vortices does not vary much with $f$.

\subsection{Maximal velocity at different frequencies}

\begin{figure}
\centering
\includegraphics[width=0.55\textwidth]{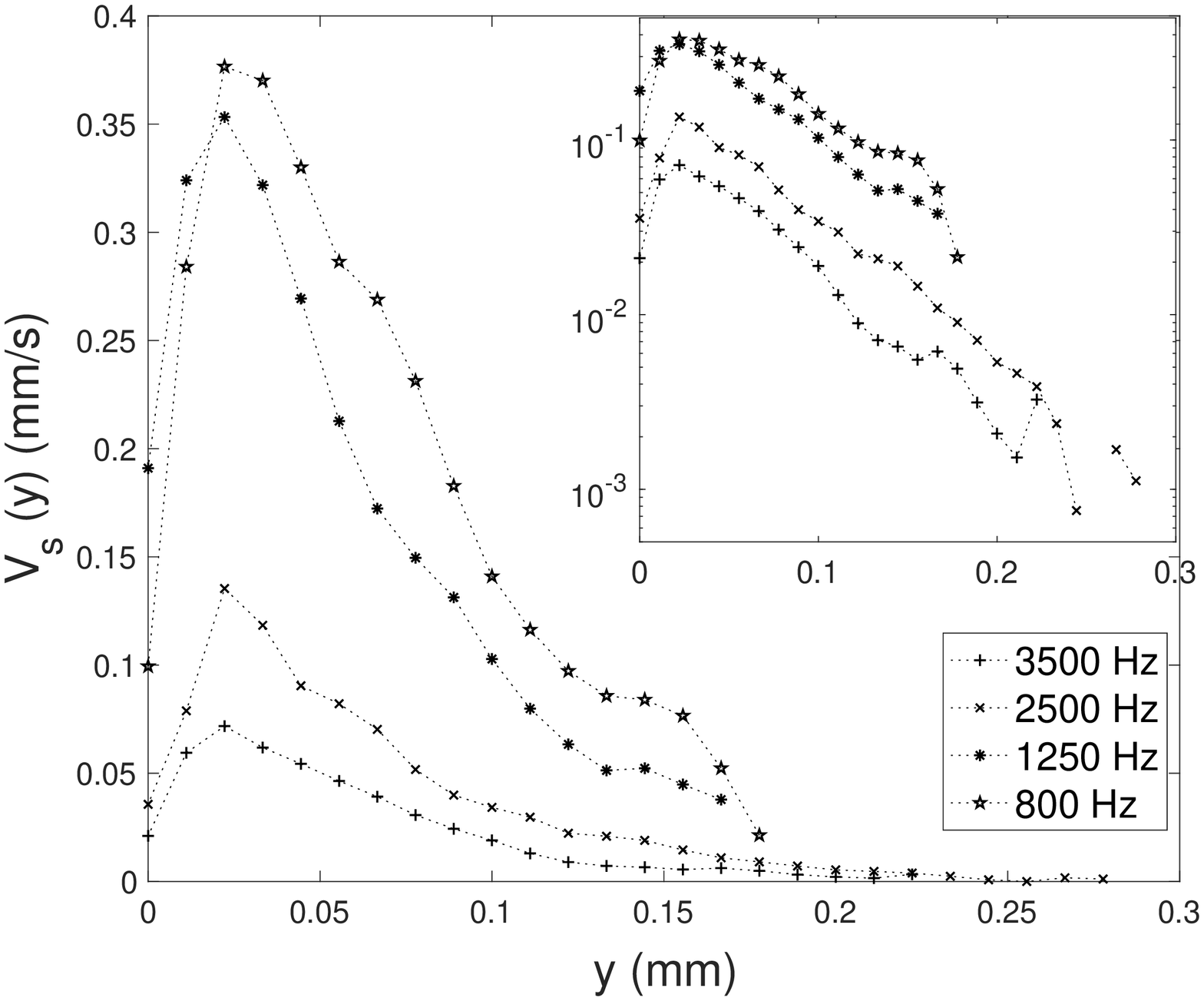}
\caption{Streaming velocity profile along vertical direction $V_s (y)$, for four different frequencies. Liquid viscosity $\nu$ = 4.32 mm$^2$/s and $V_a$ = 22 mm/s. The inset plots the same data in Lin-Log axes.}
\label{fig:influ_4freq}
\end{figure}

\begin{figure}
\centering
\includegraphics[width=0.55\textwidth]{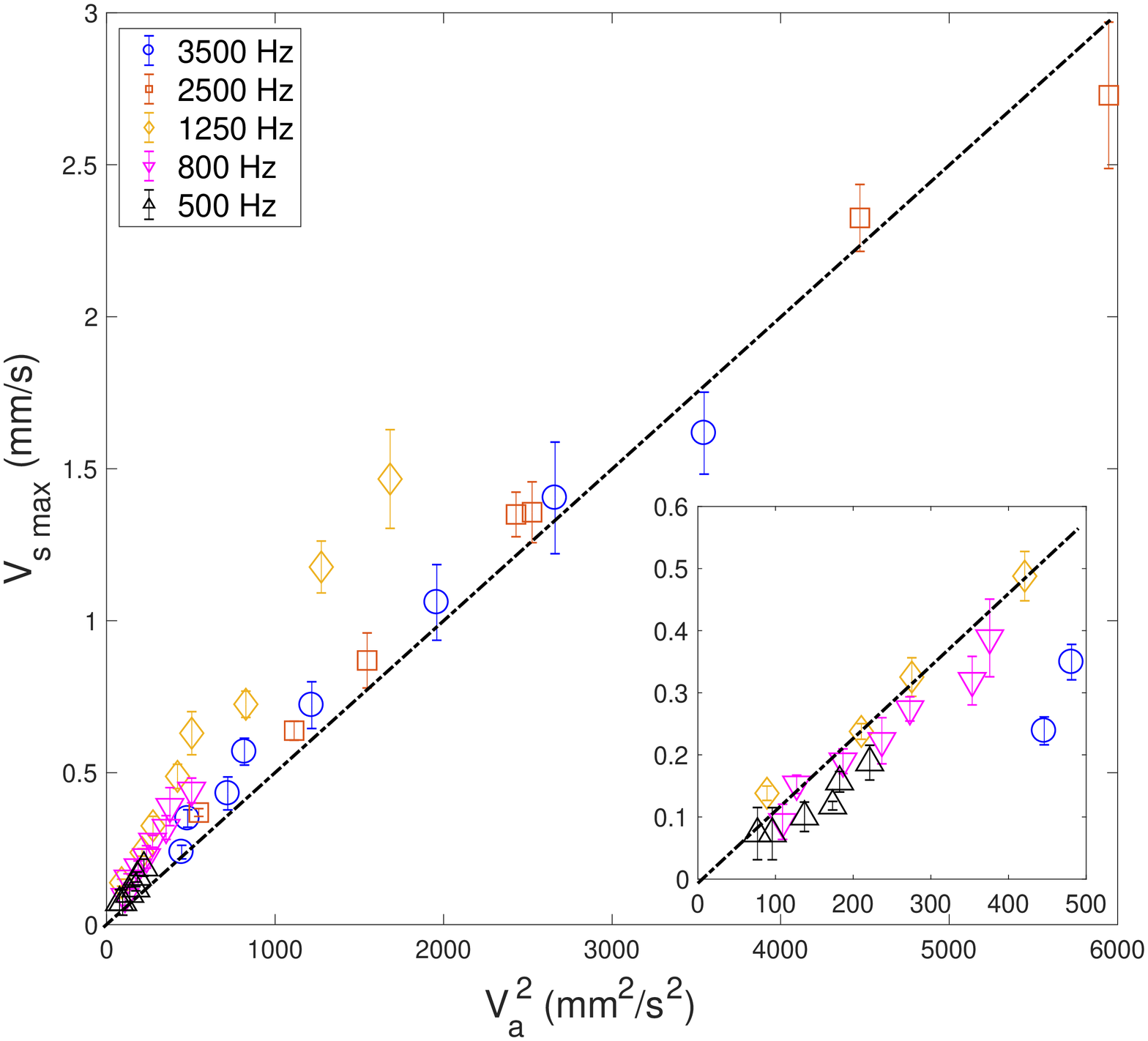}
\caption{Maximal streaming velocity $V_\text{{s\,max}}$ versus $V_a^{ 2}$, for different $f$ and the same viscosity $\nu$ = 4.32 mm$^2$/s. The dashed-dotted line suggests a linear relationship, with a prefactor $\theta$ = 5$\times$10$^{-4}$ s/mm. The inset represents a magnified view of the plot for the lowest values of $V_a^{ 2}$, suggesting a linear scaling with a prefactor $\theta$ = 0.0011 s/mm.}
\label{fig:influ_freq}
\end{figure}

We extract the velocity profile $V_s (x=0,y)$ for the four values of frequency, under the same conditions as those of Figs.~\ref{fig:influ_freq_piv} and \ref{fig:influ_freq_vort}, in particular $V_a$ is fixed at 22 mm/s. Results are plotted in Figure \ref{fig:influ_4freq}. The $y$ locations of the maxima roughly correspond to the VBL thickness at respective $f$ : $\delta_{3500} \simeq$ 19.8 $\mu$m, $\delta_{2500} \simeq$ 23.4 $\mu$m, $\delta_{1250} \simeq$ = 33.2 $\mu$m and $\delta_{800} \simeq$ 41.5 $\mu$m. The maximal velocity itself is very much dependent on $f$, but the typical length-scale of the decay along $y$ is comparable for all four experiments, as revealed by the Lin-Log plot in the insert. The four velocity profiles are shifted from each other with a given offset.

\begin{figure}
\centering
\subfigure[]{\includegraphics[width=0.48\textwidth]{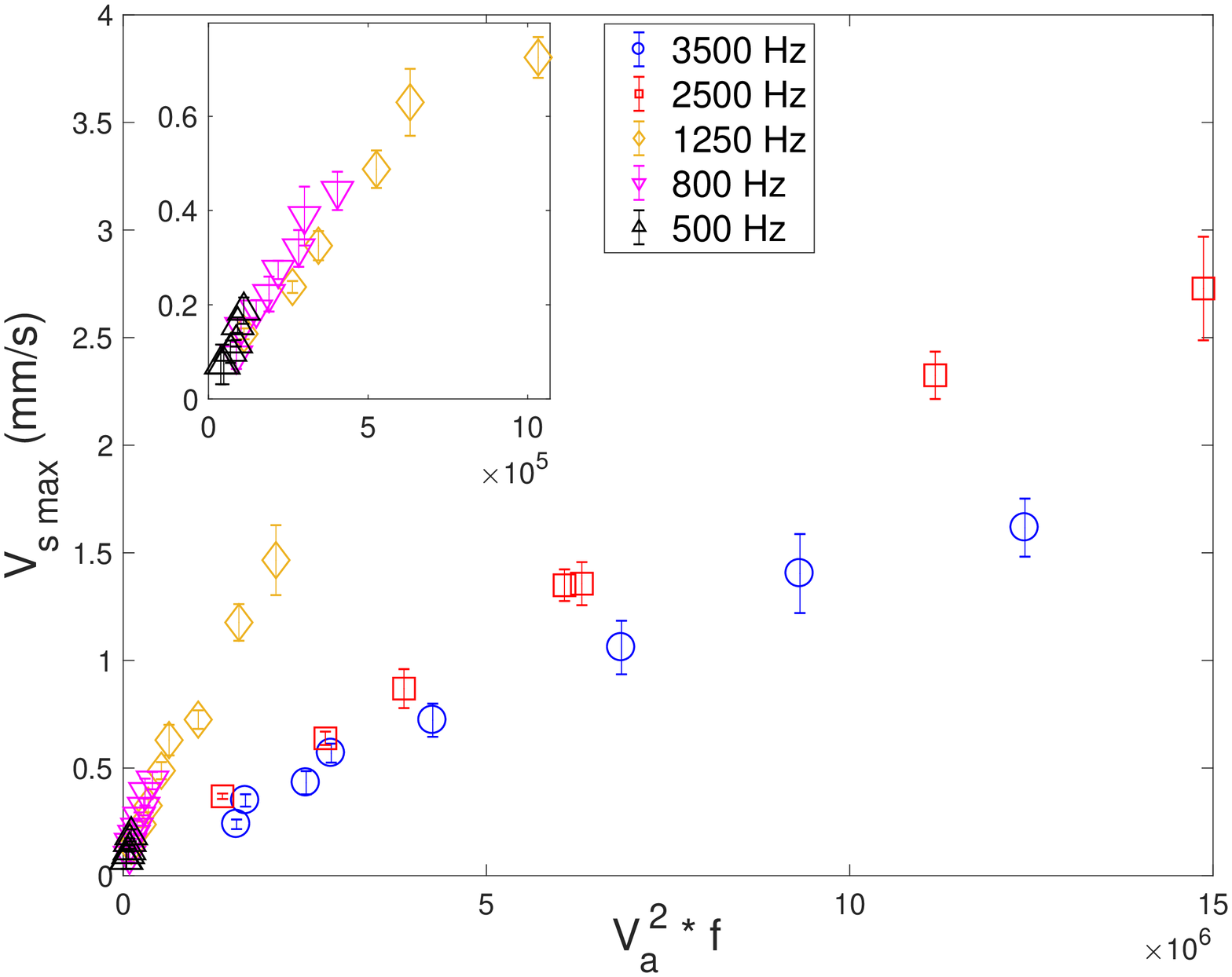}}
\subfigure[]{\includegraphics[width=0.48\textwidth]{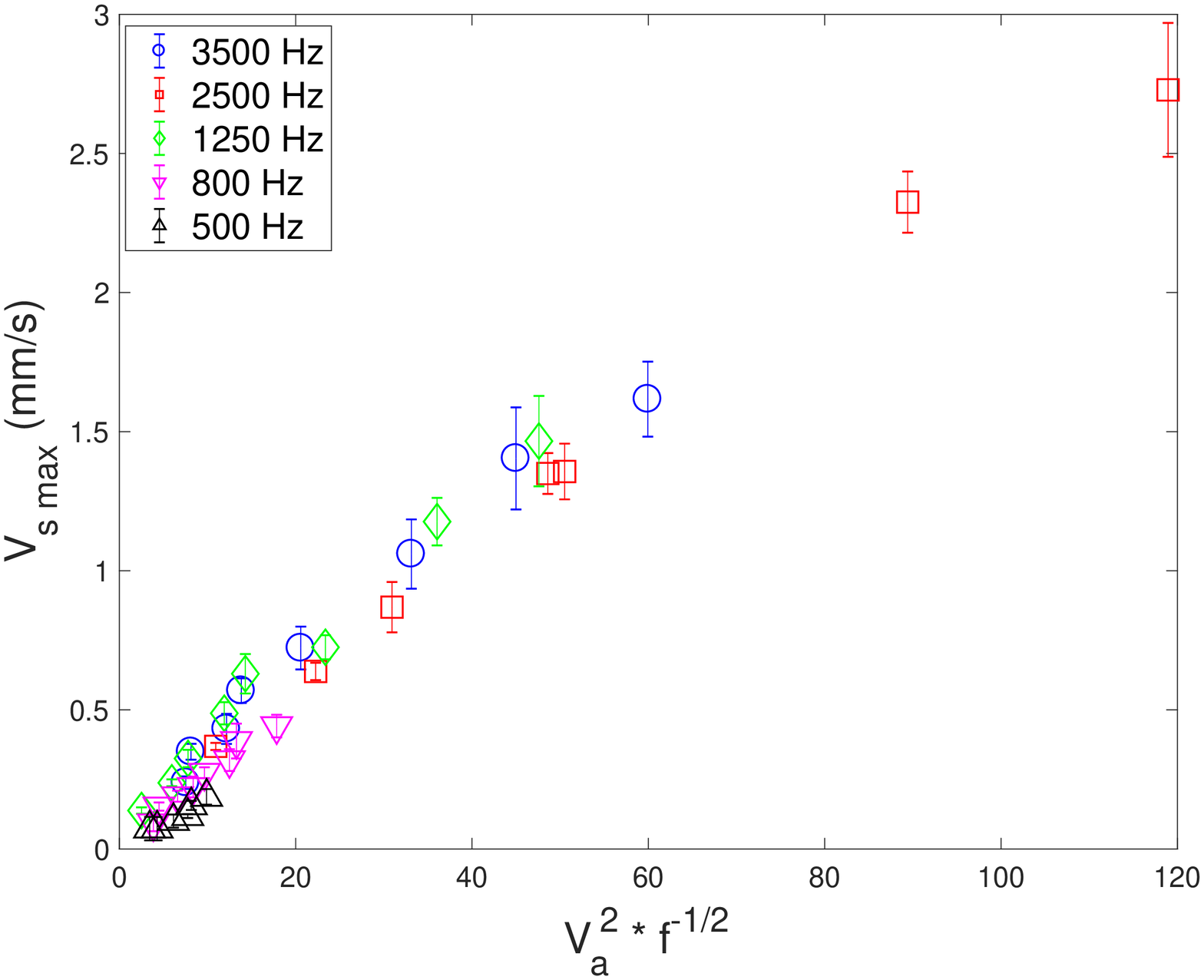}}
\caption{Attempts of data rescalling for $V_\text{{s\,max}}$ (a) versus $V_a^{ 2}\times f$ (insert shows data in the lowest range of $V_a^{ 2}$ and, (b) versus $V_a^{ 2}\times f^{-1/2}$ showing a fair collapse of data.}
\label{fig:influ_freq2}
\end{figure}

Figure \ref{fig:influ_freq} shows the maximal velocity $V_\text{{s\,max}}$ versus the square of the acoustic forcing velocity $V_a^{ 2}$, for different values of frequencies $f$ and the same liquid viscosity $\nu$ = 4.32 mm$^2$/s. Each data group obtained at constant $f$ shows a linear trend : $V_\text{{s\,max}} = \theta V_a^{ 2}$. However, the dependence of the prefactor $\theta$ on $f$ is unclear. Obviously, the theoretical prediction of \cite{Ovchinnikov2014} shown in eq.~(\ref{eq:th_predict}) fails to predict this strong dependence on $f$. However, it is possible to make two groups of data : 

- one group rather concerns measurements obtained at higher frequencies (2500 and 3500 Hz) and high $V_a$, for which a good fit is obtained for a value $\theta$ = 5$\times$10$^{-4}$ s/mm.
 
 - the other group is constituted by measurements obtained at lower frequencies (500, 800 and 1250 Hz) and relatively low $V_a$, see insert in Fig.~\ref{fig:influ_freq}. \textcolor{red}{In this case, the value of the prefactor is $\theta$ = 0.0011 s/mm.} 

To further test the possibility of a scaling law that would capture the dependence of the streaming velocity on $f$, we attempted to plot $V_{\text{s\,max}}$ versus potential pertinent combinations of powers of $V_a$ and $f$. In classical Rayleigh-Schlichting streaming, $V_{\text{s\,max}}$ usually depends linearly to $A^2 f = V_a^2 / (4 \pi^2 f) $ \cite{Costalonga2015,Bahrani2020}. But it turns out that plotting $V_{\text{s\,max}}$ versus $V_a^2 / (4 \pi^2 f) $ leads to even more scattered data points.

In the seek for an empirical law quantifying the dependence on $f$, we then tried to plot $V_s$ versus other combinations of $V_a^2$ and $f^{\beta}$, with $\beta$ being a real exponent, predicted to equal -0.1 from Ovchinnikov \textit{et al.}'s theory \cite{Ovchinnikov2014}, see eq.~(\ref{eq:th_predict}). Figures \ref{fig:influ_freq2} show the two most successful attempts :

- Figure \ref{fig:influ_freq2}-(a) : the plot of $V_s$ versus $V_a^{ 2} \times f$ shows a good collapse of data for the three lowest frequency values (500, 800 and 1250 Hz). But the rescaling does not fit with the two other data sets corresponding to the highest frequencies (2500 and 3500 Hz).

- Figure \ref{fig:influ_freq2}-(b) : the plot of $V_s$ versus $V_a^{ 2} \times f^{-1/2}$ shows a fair collapse of data for all frequencies, though it is more convincing at higher acoustic amplitude.

Still, there is no clear explanation for such trends. Therefore, it is likely that the dependence of the streaming flow on $f$ cannot be captured by simple theoretical predictions. 

\section{Conclusions}

Our study presents qualitative and quantitative results of the streaming flow generated by long-wavelength/low-frequency acoustic fields near a sharp-edge. Main attention has been given to viscosity ($\nu$ from pure water to 30 times higher), frequency $f$ from 500 to 3500 Hz, allowing to tune the VBL thickness $\delta$ from 9.5 to 137 $\mu$m. The mechanisms of such a streaming flow, described in previous studies \cite{Huang2013a,Huang2014,Nama2014,Nama2016a,Ovchinnikov2014,Zhang2019,DoinikovMN2020}, are distinct from those of the classical Rayleigh-Schlichting streaming. Our results confirm a strong link of sharp-edge streaming with viscosity and frequency. But the dependency on both $\nu$ and $f$ seems to be more complex than simple power law descriptions, like for instance those from Ovchinnikov {et al.}'s study \cite{Ovchinnikov2014}. \textcolor{red}{Let us mention a very recent study \cite{DoinikovMN2020} where streaming velocity is predicted analytically and numerically. Equations (27-28) and (37-38) in \cite{DoinikovMN2020} offer a complete prediction, including the structure of the flow itself. By comparing the scaling laws from this study with our experiments, we could not find agreement. We assume the complex behavior in our experiments is due to that $\delta$ can become comparable to the channel depth. Therefore, we hope our results will give interesting challenge for future studies involving complex geometries.}.

Still, our results allow to draw several conclusions :

- for any conditions, the maximal streaming velocity is roughly located at a vertical distance of $\delta$ from the tip, i.e. just at the limit of the VBL.

- an increase of viscosity leads to globally weaken the streaming velocity and the outer vorticity. Still, the outer vortices keep their size and shape for all liquids, and the thickness of the inner flow along the edge lateral walls roughly remain insensitive to viscosity. This is clearly at odds from what is observed in classical boundary-layer (Rayleigh-Schlichting) streaming.

- at constant $V_a$, a decrease of frequency tends to increase the streaming velocity. Our results, although unexplained by the current theoretical state of the art, suggests the empirical law: $V_s \sim V_a^{ 2} f^{-1/2}$. Furthermore, the lower the frequency $f$ is, the more spread out the streaming vortices are.

- while the flow near the tip ($r < \delta$) is strongly influenced by $\nu$ and $f$, the flow far from the tip follows an exponential decrease over a length scale of roughly 130 $\mu$m, under the test condition and with angle of 60$^{\circ}$, and tip height $h$= 180 $\mu$m. This length characterises the disturbance distance and seems to be dependent only on the sharp edge structure rather than the operating conditions.

- when the VBL thickness is comparable to the channel depth, i.e. when $p^*$ of the order one, the dependence of $V_{s max}$ on $V_a^{ 2}$ is no longer linear. It suggests that $p^* \gg$ 1 is a necessary condition for this linearity, as otherwise the streaming flow cannot fully develop within the channel.


\vspace{6pt} 



\authorcontributions{X.G., L.R. and P.B. planned the work. C.Z. and P.B. fabricated the device. C.Z. carried out the experiments. C.Z., X.G., L.R. and P.B. discussed the results and their presentation in figures. P.B. wrote the first draft of the paper. C.Z., X.G., L.R. and P.B. wrote the paper. All authors have read and agreed to the published version of the manuscript.}

\funding{C.Z. was funded by the China Scholarship Council.}


\conflictsofinterest{The authors declare no conflict of interest.} 

\abbreviations{The following abbreviations are used in this manuscript:\\

\noindent 
\begin{tabular}{@{}ll}
VBL & Viscous Boundary Layer \\
\end{tabular}}

\appendixtitles{no} 
\appendix


\reftitle{References}


\externalbibliography{yes}
\bibliography{biblio_simusNS.bib}

\begin{thebibliography}{-------}
\providecommand{\natexlab}[1]{#1}

\bibitem[Westervelt(1953)]{Westervelt1953}
Westervelt, P.J.
\newblock {The Theory of Steady Rotational Flow Generated by a Sound Field}.
\newblock {\em The Journal of the Acoustical Society of America} {\bf 1953},
  {\em 25},~60--67.
\newblock
  doi:{\changeurlcolor{black}\href{https://doi.org/10.1121/1.1907009}{\detokenize{10.1121/1.1907009}}}.

\bibitem[Nyborg(1953)]{Nyborg1953}
Nyborg, W.L.
\newblock {Acoustic Streaming due to Attenuated Plane Waves}.
\newblock {\em The Journal of the Acoustical Society of America} {\bf 1953},
  {\em 25},~68--75.
\newblock
  doi:{\changeurlcolor{black}\href{https://doi.org/10.1121/1.1907010}{\detokenize{10.1121/1.1907010}}}.

\bibitem[Lighthill(1978)]{Lighthill1978}
Lighthill, S.J.
\newblock Acoustic Streaming.
\newblock {\em Journal of Sound And Vibration} {\bf 1978}, {\em 61},~391--418.

\bibitem[Friend and Yeo(2011)]{Friend2011}
Friend, J.; Yeo, L.Y.
\newblock {Microscale acoustofluidics: Microfluidics driven via acoustics and
  ultrasonics}.
\newblock {\em Reviews of Modern Physics} {\bf 2011}, {\em 83},~647.

\bibitem[Eckart(1948)]{Eckart1948}
Eckart, C.
\newblock {Vortices and streams caused by sound waves}.
\newblock {\em Physical Review} {\bf 1948}, {\em 73},~68--76.
\newblock
  doi:{\changeurlcolor{black}\href{https://doi.org/10.1103/PhysRev.73.68}{\detokenize{10.1103/PhysRev.73.68}}}.

\bibitem[Rayleigh(1884)]{Rayleigh1884}
Rayleigh, L.
\newblock On the circulation of air observed in Kundt's tubes, and on some
  allied acoustical problems.
\newblock {\em Philosophical Transactions of the Royal Society of London} {\bf
  1884}, {\em 175},~1--21.

\bibitem[Schlichting and Gersten(2017)]{Schlichting}
Schlichting, H.; Gersten, K.
\newblock {\em Boundary-Layer Theory}; Springer Nature,  2017.

\bibitem[Nyborg(1958)]{Nyborg1958}
Nyborg, W.L.
\newblock {Acoustic Streaming near a Boundary}.
\newblock {\em The Journal of the Acoustical Society of America} {\bf 1958},
  {\em 30},~329--339.
\newblock
  doi:{\changeurlcolor{black}\href{https://doi.org/10.1121/1.1909587}{\detokenize{10.1121/1.1909587}}}.

\bibitem[Riley(1998)]{Riley1998a}
Riley, N.
\newblock {\em {Acoustic streaming}}; Vol.~10, Springer US: Boston, MA,  1998;
  pp. 349--356.
\newblock
  doi:{\changeurlcolor{black}\href{https://doi.org/10.1007/s001620050068}{\detokenize{10.1007/s001620050068}}}.

\bibitem[Rayleigh(1945)]{Rayleigh2013}
Rayleigh, L.
\newblock {\em {The Theory of Sound, Volume One.}}; Dover Publications,  1945;
  p. 985.

\bibitem[Faraday(1831)]{Faraday1831}
Faraday, M.
\newblock {On a Peculiar Class of Acoustical Figures; and on Certain Forms
  Assumed by Groups of Particles upon Vibrating Elastic Surfaces}.
\newblock {\em Philosophical Transactions of the Royal Society of London} {\bf
  1831}, {\em 121},~299--340.
\newblock
  doi:{\changeurlcolor{black}\href{https://doi.org/10.1098/rstl.1831.0018}{\detokenize{10.1098/rstl.1831.0018}}}.

\bibitem[Sritharan \em{et~al.}(2006)Sritharan, Strobl, Schneider, and
  Wixforth]{Sritharan2006}
Sritharan, K.; Strobl, C.J.; Schneider, M.F.; Wixforth, A.
\newblock Acoustic mixing at low Reynold's numbers.
\newblock {\em Applied Physics Letters} {\bf 2006}, {\em 88},~054102.

\bibitem[Franke \em{et~al.}(2010)Franke, Braunmuller, Schmid, Wixforth, and
  Weitz]{Franke2010}
Franke, T.; Braunmuller, S.; Schmid, L.; Wixforth, A.; Weitz, D.A.
\newblock {Surface acoustic wave actuated cell sorting (SAWACS)}.
\newblock {\em Lab on a Chip} {\bf 2010}, {\em 10},~789--794.

\bibitem[Lenshof \em{et~al.}(2012)Lenshof, Magnusson, and Laurell]{Lenshof2012}
Lenshof, A.; Magnusson, C.; Laurell, T.
\newblock {Acoustofluidics 8: Applications ofacoustophoresis in continuous
  flowmicrosystems}.
\newblock {\em Lab on a Chip} {\bf 2012}, {\em 12},~1210.

\bibitem[Sadhal(2012)]{Sadhal2012a}
Sadhal, S.S.
\newblock {Acoustofluidics 15: streaming with sound waves interacting with
  solid particles}.
\newblock {\em Lab on a Chip} {\bf 2012}, {\em 12},~2600.
\newblock
  doi:{\changeurlcolor{black}\href{https://doi.org/10.1039/c2lc40243b}{\detokenize{10.1039/c2lc40243b}}}.

\bibitem[Muller \em{et~al.}(2013)Muller, Rossi, Marin, Barnkop, Augustsson,
  Laurell, Kahler, and Bruus]{Muller2013}
Muller, P.B.; Rossi, M.; Marin, A.G.; Barnkop, R.; Augustsson, P.; Laurell, T.;
  Kahler, C.J.; Bruus, H.
\newblock {Ultrasound-induced acoustophoretic motion of microparticles in three
  dimensions}.
\newblock {\em Physical Review E} {\bf 2013}, {\em 88},~023006.

\bibitem[Skov \em{et~al.}(2019)Skov, Sehgal, Kirby, and Bruus]{Skov2019}
Skov, N.R.; Sehgal, P.; Kirby, B.J.; Bruus, H.
\newblock {Three-Dimensional Numerical Modeling of Surface-Acoustic-Wave
  Devices: Acoustophoresis of Micro-and Nanoparticles Including Streaming}.
\newblock {\em Physical Review Applied} {\bf 2019}, {\em 12},~044028.
\newblock
  doi:{\changeurlcolor{black}\href{https://doi.org/10.1103/PhysRevApplied.12.044028}{\detokenize{10.1103/PhysRevApplied.12.044028}}}.

\bibitem[Qiu \em{et~al.}(2019)Qiu, Karlsen, Bruus, and Augustsson]{Qiu2019}
Qiu, W.; Karlsen, J.T.; Bruus, H.; Augustsson, P.
\newblock {Experimental Characterization of Acoustic Streaming in Gradients of
  Density and Compressibility}.
\newblock {\em Physical Review Applied} {\bf 2019}, {\em 11},~024018.
\newblock
  doi:{\changeurlcolor{black}\href{https://doi.org/10.1103/PhysRevApplied.11.024018}{\detokenize{10.1103/PhysRevApplied.11.024018}}}.

\bibitem[Voth \em{et~al.}(2002)Voth, Bigger, Buckley, Losert, Brenner, Stone,
  and Gollub]{Voth2002}
Voth, G.A.; Bigger, B.; Buckley, M.R.; Losert, W.; Brenner, M.P.; Stone, H.A.;
  Gollub, J.P.
\newblock {Ordered clusters and dynamical states of particles in a vibrated
  fluid}.
\newblock {\em Physical Review Letters} {\bf 2002}, {\em 88},~234301.

\bibitem[Vuillermet \em{et~al.}(2016)Vuillermet, Gires, Casset, and
  Poulain]{Vuillermet2016}
Vuillermet, G.; Gires, P.Y.; Casset, F.; Poulain, C.
\newblock Chladni Patterns in a Liquid at Microscale.
\newblock {\em Physical Review Letters} {\bf 2016}, {\em 116},~184501.

\bibitem[Legay \em{et~al.}(2012)Legay, Simony, Boldo, Gondrexon, {Le Person},
  and Bontemps]{Legay2012}
Legay, M.; Simony, B.; Boldo, P.; Gondrexon, N.; {Le Person}, S.; Bontemps, A.
\newblock {Improvement of heat transfer by means of ultrasound: Application to
  a double-tube heat exchanger}.
\newblock {\em Ultrasonics Sonochemistry} {\bf 2012}, {\em 19},~1194--1200.
\newblock
  doi:{\changeurlcolor{black}\href{https://doi.org/10.1016/J.ULTSONCH.2012.04.001}{\detokenize{10.1016/J.ULTSONCH.2012.04.001}}}.

\bibitem[Loh \em{et~al.}(2002)Loh, Hyun, Ro, and Kleinstreuer]{Loh2002}
Loh, B.G.; Hyun, S.; Ro, P.I.; Kleinstreuer, C.
\newblock Acoustic streaming induced by ultrasonic flexural vibrations and
  associated enhancement of convective heat transfer.
\newblock {\em Journal of Acoustical Society of America} {\bf 2002}, {\em
  111},~875--883.

\bibitem[Kamakura \em{et~al.}(1996)Kamakura, Sudo, Matsuda, and
  Kumamoto]{Kamakura1996}
Kamakura, T.; Sudo, T.; Matsuda, K.; Kumamoto, Y.
\newblock Time evolution of acoustic streaming from a planar ultrasound source.
\newblock {\em The Journal of the Acoustical Society of America} {\bf 1996},
  {\em 100},~132--138.

\bibitem[Brunet \em{et~al.}(2010)Brunet, Baudoin, Bou~Matar, and
  Zoueshtiagh]{Brunet2010}
Brunet, P.; Baudoin, M.; Bou~Matar, O.; Zoueshtiagh, F.
\newblock Droplet displacements and oscillations induced by ultrasonic surface
  acoustic waves: A quantitative study.
\newblock {\em Physical Review E} {\bf 2010}, {\em 81},~036315.

\bibitem[Moudjed \em{et~al.}(2014)Moudjed, Botton, Henry, Ben~Hadid, and
  Garandet]{Moudjed2014}
Moudjed, B.; Botton, V.; Henry, D.; Ben~Hadid, H.; Garandet, J.P.
\newblock Scaling and dimensional analysis of acoustic streaming jets.
\newblock {\em Physics of Fluids} {\bf 2014}, {\em 26},~093602.

\bibitem[Da~Costa~Andrade(1931)]{Andrade1931}
Da~Costa~Andrade, E.N.
\newblock On the circulations caused by the vibration of air in a tube.
\newblock {\em Proceedings of the Royal Society A} {\bf 1931}, {\em 134},~445.

\bibitem[Valverde(2015)]{Valverde2015}
Valverde, J.M.
\newblock Pattern-formation under acoustic driving forces.
\newblock {\em Contemporary Physics} {\bf 2015}, {\em 56},~338--358.

\bibitem[Hamilton \em{et~al.}(2002)Hamilton, Ilinskii, and
  Zabolotskaya]{Hamilton2002}
Hamilton, M.F.; Ilinskii, Y.A.; Zabolotskaya, E.
\newblock Acoustic streaming generated by standing waves in two-dimensional
  channels of arbitrary width.
\newblock {\em Journal of Acoustical Society of America} {\bf 2002}, {\em
  113},~153--160.

\bibitem[Wiklund \em{et~al.}(2012)Wiklund, Green, and Ohlin]{Wiklund2012a}
Wiklund, M.; Green, R.; Ohlin, M.
\newblock {Acoustofluidics 14: Applications of acoustic streaming in
  microfluidic devices}.
\newblock {\em Lab on a Chip} {\bf 2012}, {\em 12},~2438.
\newblock
  doi:{\changeurlcolor{black}\href{https://doi.org/10.1039/c2lc40203c}{\detokenize{10.1039/c2lc40203c}}}.

\bibitem[Huang \em{et~al.}(2013)Huang, Xie, Ahmed, Rufo, Nama, Chen, Chan, and
  Huang]{Huang2013a}
Huang, P.H.; Xie, Y.; Ahmed, D.; Rufo, J.; Nama, N.; Chen, Y.; Chan, C.Y.;
  Huang, T.J.
\newblock {An acoustofluidic micromixer based on oscillating sidewall
  sharp-edges}.
\newblock {\em Lab on a Chip} {\bf 2013}, {\em 13},~3847--3852.
\newblock
  doi:{\changeurlcolor{black}\href{https://doi.org/10.1039/c3lc50568e}{\detokenize{10.1039/c3lc50568e}}}.

\bibitem[Huang \em{et~al.}(2014)Huang, Nama, Mao, Li, Rufo, Chen, Xie, Wei,
  Wang, and Huang]{Huang2014}
Huang, P.H.; Nama, N.; Mao, Z.; Li, P.; Rufo, J.; Chen, Y.; Xie, Y.; Wei, C.H.;
  Wang, L.; Huang, T.J.
\newblock {A reliable and programmable acoustofluidic pump powered by
  oscillating sharp-edge structures}.
\newblock {\em Lab on a Chip} {\bf 2014}, {\em 14},~4319--4323.

\bibitem[Nama \em{et~al.}(2014)Nama, Huang, Huang, and Costanzo]{Nama2014}
Nama, N.; Huang, P.H.; Huang, T.J.; Costanzo, F.
\newblock {Investigation of acoustic streaming patterns around oscillating
  sharp edges}.
\newblock {\em Lab on a Chip} {\bf 2014}, {\em 14},~2824--2836.

\bibitem[Nama \em{et~al.}(2016)Nama, Huang, Huang, and Costanzo]{Nama2016a}
Nama, N.; Huang, P.H.; Huang, T.J.; Costanzo, F.
\newblock {Investigation of micromixing by acoustically oscillated
  sharp-edges}.
\newblock {\em Biomicrofluidics} {\bf 2016}, {\em 10},~024124.

\bibitem[Doinikov \em{et~al.}(2020)Doinikov, Gerlt, Pavlic, and
  Dual]{DoinikovMN2020}
Doinikov, A.A.; Gerlt, M.S.; Pavlic, A.; Dual, J.
\newblock Acoustic streaming produced by sharp‑edge structures in
  microfluidic devices.
\newblock {\em Microfluidics and Nanofluidics} {\bf 2020}, {\em 24},~32.

\bibitem[Zhang \em{et~al.}(2019)Zhang, Guo, Brunet, Costalonga, and
  Royon]{Zhang2019}
Zhang, C.; Guo, X.; Brunet, P.; Costalonga, M.; Royon, L.
\newblock {Acoustic streaming near a sharp structure and its mixing performance
  characterization}.
\newblock {\em Microfluidics and Nanofluidics} {\bf 2019}, {\em 23},~104.
\newblock
  doi:{\changeurlcolor{black}\href{https://doi.org/10.1007/s10404-019-2271-5}{\detokenize{10.1007/s10404-019-2271-5}}}.

\bibitem[Zhang \em{et~al.}(2020)Zhang, Guo, Royon, and Brunet]{Zhang2020}
Zhang, C.; Guo, X.; Royon, L.; Brunet, P.
\newblock Unveiling of the mechanisms of acoustic streaming induced by sharp
  edges.
\newblock {\em arXiv:2003.01208} {\bf 2020}.

\bibitem[Ovchinnikov \em{et~al.}(2014)Ovchinnikov, Zhou, and
  Yalamanchili]{Ovchinnikov2014}
Ovchinnikov, M.; Zhou, J.; Yalamanchili, S.
\newblock {Acoustic streaming of a sharp edge}.
\newblock {\em The Journal of the Acoustical Society of America} {\bf 2014},
  {\em 136},~22--29.
\newblock
  doi:{\changeurlcolor{black}\href{https://doi.org/10.1121/1.4881919}{\detokenize{10.1121/1.4881919}}}.

\bibitem[Huang \em{et~al.}(2018)Huang, Chan, Li, Wang, Nama, Bachman, and
  Huang]{Huang2018a}
Huang, P.H.; Chan, C.Y.; Li, P.; Wang, Y.; Nama, N.; Bachman, H.; Huang, T.J.
\newblock {A sharp-edge-based acoustofluidic chemical signal generator}.
\newblock {\em Lab on a Chip} {\bf 2018}, {\em 18},~1411--1421.
\newblock
  doi:{\changeurlcolor{black}\href{https://doi.org/10.1039/C8LC00193F}{\detokenize{10.1039/C8LC00193F}}}.

\bibitem[Leibacher \em{et~al.}(2015)Leibacher, Hahn, and Dual]{Leibacher2015}
Leibacher, I.; Hahn, P.; Dual, J.
\newblock {Acoustophoretic cell and particle trapping on microfluidic sharp
  edges}.
\newblock {\em Microfluidics and Nanofluidics} {\bf 2015}, {\em 19},~923--933.
\newblock
  doi:{\changeurlcolor{black}\href{https://doi.org/10.1007/s10404-015-1621-1}{\detokenize{10.1007/s10404-015-1621-1}}}.

\bibitem[Cao and Lu(2016)]{Cao2016}
Cao, Z.; Lu, C.
\newblock {A Microfluidic Device with Integrated Sonication and
  Immunoprecipitation for Sensitive Epigenetic Assays}.
\newblock {\em Analytical Chemistry} {\bf 2016}, {\em 88},~1965--1972.
\newblock
  doi:{\changeurlcolor{black}\href{https://doi.org/10.1021/acs.analchem.5b04707}{\detokenize{10.1021/acs.analchem.5b04707}}}.

\bibitem[Bachman \em{et~al.}(2018)Bachman, Huang, Zhao, Yang, Zhang, Fu, and
  Huang]{Bachman2018}
Bachman, H.; Huang, P.H.; Zhao, S.; Yang, S.; Zhang, P.; Fu, H.; Huang, T.J.
\newblock {Acoustofluidic devices controlled by cell phones}.
\newblock {\em Lab on a Chip} {\bf 2018}, {\em 18},~433--441.
\newblock
  doi:{\changeurlcolor{black}\href{https://doi.org/10.1039/C7LC01222E}{\detokenize{10.1039/C7LC01222E}}}.

\bibitem[Costalonga \em{et~al.}(2015)Costalonga, Brunet, and
  Peerhossaini]{Costalonga2015}
Costalonga, M.; Brunet, P.; Peerhossaini, H.
\newblock {Low frequency vibration induced streaming in a Hele-Shaw cell}.
\newblock {\em Physics of Fluids} {\bf 2015}, {\em 27},~013101.

\bibitem[Cheng(2008)]{Cheng2008}
Cheng, N.S.
\newblock Formula for the viscosity of a glycerol-water mixture.
\newblock {\em Ind. Engng Chem. Res.} {\bf 2008}, {\em 47},~3285--3288.

\bibitem[Slie \em{et~al.}(1966)Slie, Donfor, and Litovitz]{Slie1966}
Slie, W.M.; Donfor, A.R.; Litovitz, T.A.
\newblock Ultrasonic shear and longitudinal measurements in aqueous glycerol.
\newblock {\em Journal of Chemical Physics} {\bf 1966}, {\em 44},~3712--3718.

\bibitem[Bahrani \em{et~al.}(2020)Bahrani, Perinet, Costalonga, Royon, and
  Brunet]{Bahrani2020}
Bahrani, S.; Perinet, N.; Costalonga, M.; Royon, L.; Brunet, P.
\newblock Vortex elongation in outer streaming flows.
\newblock {\em Experiments in Fluids} {\bf 2020}, {\em 61},~91.

\end{thebibliography}





\end{document}